\documentclass[aps,11pt,showkeys,nofootinbib]{revtex4-1}

\usepackage{amsmath,amssymb,amsfonts,amsthm}
\usepackage{amsbsy} 
\usepackage{epsfig}
\usepackage{latexsym}

\usepackage[dvipsnames]{xcolor}
\usepackage{soul} 
\usepackage{subcaption}
\usepackage{braket}
\usepackage{tensor}
\usepackage{ulem}

\usepackage{hyperref}

\newcommand{\mk}[0]{\mathcal{K}}

\newcommand{\ma}[0]{\mathcal{A}}
\newcommand{\mb}[0]{\mathcal{B}}

\newcommand{\Laplace}{\Delta}

\newcommand{\sgn}[1]{\text{sgn}(#1)}

\newcommand{\mkz}{\mathcal{K}^{(0)}}
\newcommand{\mkt}{\mathcal{K}^{(2)}}

\newcommand{\chiz}{(\chi^0)}

\newcommand{\ev}[1]{\bra{\sigma} #1 \ket{\sigma}}

\newcommand{\piInt}[1]{\int \frac{{\rm d}^3 #1}{(2 \pi)^3}}

\newcommand{\pd}{\partial}
\newcommand{\ii}{\text{i}}
\newcommand{\Tmat}{^{(\chi)}\! T}

\newcommand{\no}[1]{:#1\!:}

\def\be{\begin{equation}}
\def\ee{\end{equation}}
\def\dd{{\rm d}}
\def\bes{\begin{eqnarray}}
\def\ees{\end{eqnarray}}
\def\T{{\mathcal{T}}}

\begin{document}

\title{Reconstructing the metric in group field theory
}
\author{Steffen Gielen}
\email{s.c.gielen@sheffield.ac.uk}
\author{Lisa Mickel}
\email{lmickel1@sheffield.ac.uk}
\affiliation{School of Mathematics and Statistics, University of Sheffield, Hicks Building, Hounsfield Road, Sheffield S3 7RH, United Kingdom}
\date{\today}

\begin{abstract}
We study a group field theory (GFT) for quantum gravity coupled to four massless scalar fields, using these matter fields to define a (relational) coordinate system. We exploit symmetries of the GFT action, in particular under shifts in the values of the scalar fields, to derive a set of classically conserved currents, and show that the same conservation laws hold exactly at the quantum level regardless of the choice of state. 
We propose a natural interpretation of the conserved currents which implies that the matter fields always satisfy the Klein--Gordon equation in GFT. 
We then observe that in our matter reference frame, the same conserved currents can be used to extract all components of an effective GFT spacetime metric. 
Finally, we apply this construction to the simple example of a spatially flat homogeneous and isotropic universe, where we derive an effective Friedmann equation directly from this metric. The Friedmann equation displays a bounce and a late-time limit equivalent to general relativity with a single scalar field.
Our proposal goes substantially beyond the GFT literature in which only specific geometric quantities such as the total volume or volume perturbations could be defined, opening up the possibility to study more general geometries as emerging from GFT.
\end{abstract}


\maketitle

\section{Introduction}

The conceptual foundation of classical general relativity starts from the 
notion of a spacetime metric, from which all relevant geometric properties of spacetime, as well as physical effects related to gravity, can be derived. While the dynamical equations of general relativity are formulated in diffeomorphism-covariant terms -- they take the same form no matter what coordinate system is used -- the tensorial quantities in them, most prominently the metric itself, are not diffeomorphism-invariant and therefore depend on the coordinate system. This means that the metric or curvature tensors cannot directly be observable. Constructing interesting (diffeomorphism-invariant) observables from the metric is in general a highly nontrivial task \cite{Rovelli:1990ph,Giddings:2005id,Dittrich:2005kc,Tambornino:2011vg}.

The implementation of diffeomorphism symmetry at the quantum level is often seen as one of the most formidable obstacles in the construction of a full theory of quantum gravity. Discrete approaches somewhat circumvent this problem since they no longer work with a differentiable manifold on which diffeomorphisms act, but directly with quantities such as lengths, areas or finite parallel transports which are to an extent diffeomorphism-invariant. However, such approaches then face at least two important basic challenges: one is the recovery of a continuum limit in which differentiable structures, and with them the usual freedom to choose coordinates emerge \cite{Bahr:2009ku,Dittrich:2014ala}; the other is the extraction of relevant observables, given that there is no useful way of directly defining tensorial objects, such as curvature invariants, in the discrete setting (see, e.g., \cite{Loll:2019rdj} for discussion in the setting of causal dynamical triangulations).

A common strategy to construct useful observables is to focus on relational observables \cite{Tambornino:2011vg}, particularly those built from using matter fields as coordinates. A prime example of this is homogeneous cosmology, where a massless scalar field can serve as a good clock, and the expansion of the universe can be characterised by stating the evolution of the scale factor relative to the value of the scalar field \cite{Ashtekar:2011ni}. Such a characterisation is indeed invariant under time reparametrisations. More generally, suitably chosen matter fields allow the construction of a relational coordinate system, such that the coordinates are now physical degrees of freedom rather than arbitrary gauge structure. This idea has been employed particularly in dust models (see \cite{Brown:1994py,Giesel:2007wi,Husain:2011tm,Husain:2011tk} for some of the vast literature) and gravity coupled to scalar fields \cite{Domagala:2010bm, Giesel_2019}.

In this paper we focus on the group field theory (GFT) approach to quantum gravity \cite{Freidel:2005qe,Oriti:2011jm,Oriti:2013aqa}, which is closely related to loop quantum gravity \cite{Ashtekar_2004}, matrix and tensor models \cite{Gurau_2012}. GFT is a fundamentally discrete (and background-independent) setting for quantum gravity; one does not work with fields on a manifold but with combinatorial structures from which spacetime, and all continuum matter fields, are supposed to emerge in a continuum limit. Because of this, GFT shares with other discrete approaches the issues of defining observables and in particular an analogue of a spacetime metric, which would be important in order to connect to classical gravitational theories (general relativity or extensions) or any type of phenomenology. Gravitational observables that have been constructed so far in GFT are defined in analogy with simple geometric operators in loop quantum gravity; in particular one can define a GFT volume operator (based on \cite{Rovelli:1994ge,Ashtekar:1997fb}), and from this a relational volume observable representing the total spatial volume at a given instant of relational time, here again given by a massless scalar field. This observable has been used to derive an effective Friedmann equation \cite{Oriti:2016qtz,Oriti:2016ueo}, very similar to the one in loop quantum cosmology \cite{Ashtekar:2011ni}. By coupling additional massless scalar fields that are used as ``rods'' or spatial coordinates, one can turn this global volume into a local volume element (now dependent on the spatial coordinates as well) and try and use this to define an effective cosmological perturbation theory in GFT \cite{Gielen:2017eco,Gerhardt:2018byq,Gielen:2018fqv,Marchetti:2021gcv,Jercher:2023nxa}. Such a formalism can then be used to gain initial insights on dynamics of volume perturbations around an effective homogeneous universe characterised by the Friedmann equation obtained before, but given that volume perturbations are not gauge-invariant, translating the results into the usual textbook discussions in terms of gauge-invariant quantities requires some effort. There is also no direct way of accessing tensor fluctuations which do not affect the local volume element.

In the following we will propose a new approach towards tackling this issue, based on the idea of locally conserved currents associated to GFT symmetries. All previous work on GFT models for quantum gravity with massless scalar fields, starting from \cite{Oriti:2016qtz,Oriti:2016ueo}, starts by identifying symmetries of the corresponding classical theory, and requiring that these are represented as symmetries in GFT.  In the case of cosmological models based on a single matter field, the most important symmetry is with respect to shifts in the field; this symmetry  of a classical free, massless scalar field justifies its use as a clock.  In this relational coordinate interpretation, it can be seen as a time-translation symmetry. The conserved quantity associated to this symmetry is the momentum conjugate to the scalar field, whose conservation gives the Klein--Gordon equation in the homogeneous approximation. Hence, the matter dynamics obtained from GFT are consistent with classical expectations, and the scalar field is a good clock also in GFT.

Our proposal is to extend this line of argument to four free, massless scalar fields, now used as coordinates for space and time. There are now four independent translational symmetries, leading to four conserved currents which form the analogue of the energy-momentum tensor in standard quantum field theory. At the classical level, each massless scalar field has its own Klein--Gordon current, whose conservation gives the classical field equations. 
By showing that the GFT energy-momentum tensor arising from the translational invariance of the GFT action with respect to the relational fields  is conserved classically and quantum-mechanically, we immediately obtain the Klein--Gordon equations for all matter fields, now no longer restricted to the spatially homogeneous setting. Thus, as a first major insight we show that these matter fields always satisfy the same dynamical equations in GFT as they do in standard spacetime field theory. Within the approximations we use regarding the dynamics of GFT, this is true for any state, and does not require any semiclassical approximations as are often used in the literature. We then use the fact that classically the Klein--Gordon currents depend explicitly on the metric, and so in a matter reference frame can be used to read off all components of the metric. 

Given that the GFT energy-momentum tensor represents the same physical quantities, we show how to use the energy-momentum tensor to define a spacetime metric in full GFT for sufficiently semiclassical states.
This result is the main achievement of our work, given the previous severe limitations in defining relational observables. We illustrate our new formalism in the case of homogeneous, isotropic cosmology, finding some familiar results regarding a Friedmann equation and bounce, but also some puzzling results regarding the role of the new spatial coordinate fields. These results deserve further attention, but our formalism also suggests applications to inhomogeneous cosmology, black holes or other spacetimes of interest, which will be explored in future work.

While our proposal is specific to the GFT setting, it only requires a symmetry of the action under shifts in the matter fields (to be used as coordinates) coupled to quantum gravity. As such, it could in principle also be applied to other background-independent approaches to extract an effective metric. Regarding GFT, we emphasise that our proposal differs from existing ideas in the literature to use either bivector/area operators \cite{Gielen:2013kla,Gielen:2013naa} or a volume operator \cite{Oriti:2016qtz,Oriti:2016ueo} to reconstruct metric information. Indeed, the properties of our effective metric can in general disagree with results derived from these other operators.

In section \ref{sec:current} we show how in general relativity the classical shift symmetry of four free massless scalar fields leads to conserved currents that encode the metric in a (relational) coordinate system given by these scalar fields. Section \ref{sec:GFT} gives a short review of the canonical quantisation of GFT, leading to the definition of an energy-momentum tensor. We explicitly show that the energy-momentum tensor is conserved as an operator. Section \ref{sec:cosmoEx} illustrates the general idea in a simple cosmological example. By choosing an appropriate coherent state, we find an effective metric corresponding to a spatially flat homogeneous and isotropic Universe. We also derive the effective Friedmann equation, which is similar to equations derived by other methods in the literature. At late times, this Friedmann equation reduces to the classical equation expected for a single scalar field only, leading to a discussion of why the other fields do not contribute.  We conclude in section \ref{sec:conc}.

\section{Spacetime metric as a conserved current}
\label{sec:current}

In standard field theory, a free massless scalar field $\chi$ on a curved background (with Lorentzian metric $g_{\mu\nu}$) can be defined in terms of the action
\be
S = -\frac{1}{2}\int {\rm d}^4 x\;\sqrt{-g}\, g^{\mu\nu} \partial_\mu\chi \partial_\nu\chi\,,
\label{scalaraction}
\ee
which is invariant under constant shifts in the field $\chi\mapsto\chi+\epsilon$, where $\epsilon$ is a constant. By Noether's theorem \cite{Noether,Noether_translation}, this symmetry implies a conservation law
\be
\partial_\mu j^\mu = 0\,,\quad j^\mu =-\sqrt{-g}\, g^{\mu\nu} \partial_\nu\chi\,.
\ee
This conservation law is of course nothing but the Klein--Gordon equation $\Box\chi=0$. Notice that $j^0$ is equal to $\pi_\chi$, the canonical momentum of $\chi$.

Let us now identify the scalar field with a spacetime coordinate $x^A$ (i.e., surfaces of constant $x^A$ are taken to be surfaces of constant $\chi$). In this case, by definition, we have $\partial_\mu\chi=\delta^A_\mu$ and hence\footnote[0]{In this arXiv version we include minor clarifications of sign choices, which do not appear in the published version, in footnotes -1 to -6.}\footnote[-1]{Notice in particular that this gauge choice requires that $(j^0)^0$, the canonical momentum of $\chi^0$, is positive.}
\be
\left(j^\mu\right)^A = -\sqrt{-g}\, g^{\mu A}\,.
\label{jcurrent}
\ee
We can use four such free massless scalar fields $\chi^A,\;A=0,\ldots,3$ to define an entire relational coordinate system by identifying each spacetime point with the values of all $\chi^A$ taken at that point. The resulting relational coordinate system is locally well-defined as long as we assume the non-degeneracy condition (with respect to an arbitrary well-defined coordinate system)\footnote{For a single scalar field to be used as clock, the equivalent condition is $\partial_t\chi\neq 0$, which is (for a free massless scalar field) almost always the case in homogeneous cosmology. Outside of spatial homogeneity and for four scalars, it is less straightforward to say in general where this condition is satisfied, but generic configurations satisfy \eqref{eq:nondeg}.} 
\be
\det(\partial_\mu\chi^A)\neq 0\,.
\label{eq:nondeg}
\ee
In this coordinate system, where the gradients of the scalar fields are mere numbers and thereby dimensionless, the metric components have units of length$^4$, $[g_{AB}] = L^4$, and the conserved current has $[(j^\mu)^A] = L^4$ (with $\hbar=c=1$). In principle, we could have inserted an arbitrary dimensionful proportionality factor $\xi$ when fixing the coordinate system, $\pd_\mu \chi^A = \xi \delta^A_\mu$, but as $\xi$ just represents a unit convention without physical significance, the simplest choice is $\xi=1$.

In this coordinate system, knowledge of the currents on the left-hand side of (\ref{jcurrent}) can be used to define a symmetric matrix field $j^{AB}=\left(j^A\right)^B$ which defines the inverse metric (in this coordinate system) as
\be
g^{AB} = -\left(-\det(j^{AB})\right)^{-1/2} j^{AB}\,.
\label{inversemetric}
\ee
The positioning of the capital Latin index on the left-hand side of \eqref{jcurrent} is initially conventional, given that it merely corresponds to a label for the different matter fields. However, once we define $j^{AB}$ and establish its relation to the inverse metric, the notation becomes more intuitive if we think of $A, \, B$ as contravariant indices.

There are two minus signs in the relation (\ref{inversemetric}); the one inside the brackets comes from the assumption that $g_{\mu\nu}$, and hence also $j^{AB}$, has negative determinant (coming from a Lorentzian signature). The overall minus sign can be traced back to the minus sign in the action (\ref{scalaraction}), which comes from the assumption that the metric $g_{\mu\nu}$ has one negative and three positive eigenvalues (the ``East Coast'' signature convention). Of course, classically these assumptions are reasonable, but in a quantum gravity setting it may not be a priori clear whether we can fix the metric signature or the number of positive and negative eigenvalues; see \cite{Alexandre:2023nmh} for a recent discussion of a general classical framework in which all possible signatures may co-exist. Adopting the ``West Coast'' signature convention throughout would change the signature of $g^{\mu\nu}$ but also add an additional minus sign in (\ref{jcurrent}), leading to a $(j^\mu)^A$ of the same signature $(+---)$. We will come back to a discussion of this point when looking at concrete examples.

When coupling these matter fields to general relativity, we obtain a diffeomorphism-invariant theory in which $g_{\mu\nu}$ becomes dynamical. The condition $\partial_\mu\chi^A=\delta^A_\mu$ is a local gauge-fixing of this gauge symmetry, which can be seen as a specific case of the {\em harmonic gauge} condition $\Box x^\mu=0$ whose use has a long history in classical general relativity \cite{fouresbruhat,fock,Wald:1984rg}. Whereas the harmonic gauge condition in general does not uniquely fix the gauge (there are many solutions to it for a general $g_{\mu\nu}$), the ``scalar field gauge'' we are adopting here does fix it completely, assuming it is well-defined  by \eqref{eq:nondeg}. For free massless scalar fields, this gauge does not determine which of the coordinate directions given by $\chi^A$ are timelike, spacelike or null, unlike for dust constructions such as \cite{Brown:1994py} which are more suited to a $(3+1)$ splitting in which spacelike and timelike directions need to be separated.

The central result of this discussion is (\ref{inversemetric}), which tells us how to compute all components of the inverse metric from the symmetric $j^{AB}$. 
Due to its direct relation to the metric, $j^{AB}$ for the scalar field action given in  \eqref{scalaraction} is symmetric in its two indices when written in the relational coordinate system. (For more general scalar field actions, this might change and the construction would not work.)
In a quantum theory in which $j^{AB}$ can be defined as an operator, one can define an effective $g^{AB}$, e.g., from expectation values of $j^{AB}$ in a semiclassical state by using (\ref{inversemetric}). In the following we will see that a GFT coupled to four scalar fields does have an operator analogue of $j^{AB}$ (given by the GFT energy-momentum tensor) and hence an effective spacetime metric can be written down unambiguously.
Clearly, this operator analogue needs to be symmetric as well, which will be the case for the GFT energy-momentum tensor we construct in section \ref{sec:GFTTAB}.

\section{Group field theory}
\label{sec:GFT}

GFT can be seen as a ``quantum field theory not {\em on}, but {\em of} spacetime''. 
The basic object in any GFT is a (typically real or complex bosonic) group field $\varphi(g_i,\chi^A)$, where the $g_i$ are elements of some group and the $\chi^A$ real valued, as discussed further below. The arguments of this field do not represent coordinates on a spacetime manifold; instead, the ``particles'' associated with excitations of this group field are seen as elementary building blocks of spacetime geometry and matter, not living in a pre-defined spacetime. Concretely, such an elementary building block is most commonly identified with a tetrahedron seen as the basic unit of simplicial geometry, or equivalently with a four-valent spin network vertex as the basic structure forming loop quantum gravity spin networks \cite{Oriti:2013aqa}. In a Fock space picture, a macroscopic geometry can then only emerge from a large number of such excitations over the initial vacuum; in particular, this is because quantum fluctuations of geometric observables are suppressed in the limit of large particle number. 

The discussions of this paper are applicable to a large class of models, but for concreteness we can choose to use ${\rm SU}(2)$ variables to represent parallel transports in the GFT discrete geometry, in analogy with the basic variables of canonical loop quantum gravity. More importantly, as in previous work \cite{Gielen:2020fgi, Marchetti:2021gcv, Jercher:2023nxa}, we include four $\mathbb{R}$-valued arguments $\chi^A$, which represent scalar matter degrees of freedom. We also restrict ourselves to the case of a real group field; generalisation to a complex field should be entirely straightforward but does not seem necessary for our purposes. We thus have $\varphi : {\rm SU}(2)^4\times \mathbb{R}^4 \to \mathbb{R}$.

As in loop quantum gravity, it is useful to expand the group field in modes associated to ${\rm SU}(2)$ representation data,
\be
\varphi(g_i,\chi^A) = \sum_J \varphi_J(\chi^A)\,D_J(g_i)\, ,
\label{eq:modeDecomp}
\ee
where $D_J(g_i)$ represent suitable combinations of Wigner $D$-matrices and $J = (\Vec{j},\, \Vec{m},\, \iota)$ is a multi-index representing ${\rm SU}(2)$ irreducible representations $\Vec{j}$, magnetic indices $\Vec{m}$, and intertwiners $\iota$ (for more details see, e.g., \cite{Assanioussi:2020hwf}). In the following, we will only need to use the existence of such an expansion, and no details of ${\rm SU}(2)$ representation theory. This means that our results immediately apply to any choice of compact group for the $g_i$ variables. 

While not many explicit constructions of GFT models for quantum gravity coupled to four scalar fields exist in the literature, there are several GFT models built on the gauge group ${\rm SU}(2)$, including a version of GFT corresponding to the Engle--Pereira--Rovelli--Livine (EPRL) spin foam model studied in \cite{Oriti:2016qtz,Oriti:2016ueo}. 
Extensions to non-compact gauge groups have been studied, e.g., in the context of the Barrett--Crane model \cite{Jercher:2021bie,Jercher:2023nxa} and introduce new subtleties due to the presence of continuous representations, but could be treated with similar methods. An interesting question in the spin foam approach (and, by extension, in GFT \cite{Jercher:2022mky}) is whether choosing a compact gauge group such as ${\rm SU}(2)$, interpreted as a restriction to spacelike tetrahedra only, impacts the possible causal structure of geometries emergent from such a theory \cite{Conrady:2010kc,Simao:2021qno}. We will comment on this point at the end of the paper.

Generally, when constructing possible actions one starting point is to demand that the GFT action is invariant under symmetries representing symmetries of the matter fields one wants to include \cite{Oriti:2016qtz,Oriti:2016ueo}. As discussed above, the spacetime scalar field action (\ref{scalaraction}) is invariant under shifts in the field $\chi$. It is also invariant under reflections $\chi\mapsto -\chi$. Demanding the same symmetries in GFT means that a GFT action local in $\chi^A$ cannot depend explicitly on the $\chi^A$, and can only include derivatives of even order. If all $\chi^A$ represent physically indistinguishable matter fields, we might also require symmetry under rotations $\chi^A\mapsto {R^A}_B \chi^B$ \cite{Gielen:2020fgi}. Assuming that the quadratic part of the action is also local in the $g_i$ (for terms with higher powers of the field a certain type of nonlocality would be expected \cite{Freidel:2005qe,Oriti:2011jm}) gives the general form 
\be
S[\varphi]= \int {\rm d}^4 \chi\left(\frac{1}{2}\sum_J\sum_{n=0}^\infty \,\mathcal{K}_J^{(2n)}\,\varphi_J(\chi^A)\Delta^n\varphi_J(\chi^A)-V(\varphi)\right)\,,
\label{GFTaction}
\ee
where the $\mathcal{K}_J^{(2n)}$ are arbitrary couplings and $\Delta = \sum_A \left( \frac{\partial}{\partial \chi^A}\right)^2$ is the Laplacian on $\mathbb{R}^4$. Here we have written the quadratic part of the action explicitly and moved all higher-order terms into the potential $V(\varphi)$, which may take very different forms depending on the model.

Given that we are assuming a local theory, the sum over higher derivatives in the quadratic part needs to terminate at some finite $n$, and indeed in most models studied in detail in the literature \cite{Lahoche:2019cxt,Pithis:2020sxm,Marchetti:2022nrf} only the terms $n=0$ and $n=1$ corresponding to a mass term and a single Laplacian appear. After an integration by parts of the term involving the Laplacian, one obtains an action that only depends on $\varphi$ and first derivatives, and is hence amenable to straightforward canonical quantisation. (A theory with higher derivatives would have to be treated with more involved methods, see, e.g., \cite{Raidal:2016wop}.)

Here we have not specified the precise form of the interaction terms making up the potential $V(\varphi)$. This form can be chosen by requiring that Feynman amplitudes of the resulting interacting GFT match those of spin foam models \cite{DePietri:1999bx,Reisenberger:2000zc} (we gave some examples above) and/or by again using symmetry arguments to write down a number of possible terms, which are constrained by renormalisability \cite{Carrozza:2016vsq}. 
In our work here, we will neglect the effect of interactions.  Our discussion of classical GFT extends to interacting models but their quantum analysis will be more involved in general. Neglecting interactions is a common assumption in applications of GFT to cosmology \cite{Oriti:2016qtz,Oriti:2016ueo,Jercher:2023nxa,Calcinari:2022iss}, since these are expected to be subdominant in the very early universe. In general, the range of applicability of this approximation will be limited \cite{Gielen_2020}.

The symmetry of (\ref{GFTaction}) under translations $\chi^A\mapsto\chi^A+\epsilon^A$ for arbitrary constant $\epsilon^A$ leads to a conserved current, the GFT energy-momentum tensor (here and below $\pd_A=\frac{\pd}{\pd \chi^A}$)
\be
T^{AB} := -\frac{\partial\mathcal{L}}{\partial (\partial_A\varphi)}\partial_B\varphi+\delta^{AB}\,\mathcal{L} = \sum_J\left(\mathcal{K}_J^{(2)}\partial_A\varphi_J\,\partial_B\varphi_J\right)+\delta^{AB}\,\mathcal{L}
\label{eq:TGFTdef}
\ee
with a Lagrangian density 
\be
\mathcal{L} = \sum_J\left(\frac{1}{2}\mathcal{K}_J^{(0)}\varphi_J^2 - \frac{1}{2}\mathcal{K}_J^{(2)}\sum_A(\partial_A\varphi_J)^2\right) - V(\varphi) \, ,
\label{GFTLagrange}
\ee
in which we now assume that only the terms $n=0$ and $n=1$ are present in (\ref{GFTaction}), and we have performed the integration by parts discarding a boundary term. In these expressions, we do not need to worry too much about the positioning of $A,B,\ldots$ indices; due to the $E(4)$ symmetry of the GFT action these can be raised and lowered with the Kronecker delta $\delta_{AB}$. In the identification of these GFT quantities with classical spacetime tensors, we need to be more careful, as discussed above and further below.

Naturally, the GFT energy-momentum tensor satisfies $\partial_A T^{AB}=0$. The fact that translations in the scalar field variables $\chi^A$ represent constant shifts in 
the fields that span the relational coordinate system
as discussed in section \ref{sec:current} now suggests identifying the conserved GFT quantity $T^{AB}$ with the classically conserved current $j^{AB}$.
This leads to an effective spacetime metric $g^{AB}$ via (\ref{inversemetric}). Quantum-mechanically, the energy-momentum tensor will be represented as an operator and so our identification would amount to defining also the current $j^{AB}$ as an operator. Since the inverse metric $g^{AB}$ is a non-polynomial function of $j^{AB}$, its definition as an operator would be much less straightforward. However, rather than directly proposing an ``effective metric operator'', we are usually interested in very semiclassical GFT states to be interpreted as macroscopic effective geometries. In such a state, we can use the expectation values of $T^{AB}$ (assuming fluctuations are small) to define an effective classical metric $g^{AB}$. This is analogous to the treatment of relational volume observables in most previous work on GFT cosmology and only justified for states with suitable semiclassical properties (see, e.g., \cite{Gielen_2020}).

In the following, we construct the operators corresponding to $T^{AB}$. We then show a concrete example of a reconstruction of an effective metric from the identification with the expectation values of the GFT energy-momentum tensor in section \ref{sec:cosmoEx}, where we also comment further on the required semiclassical properties of the state.

\subsection{Canonical quantisation}

We can now implement a canonical quantisation procedure for a theory defined by (\ref{GFTLagrange}). This ``deparametrised'' approach to quantisation, in which a scalar field is used as a clock from the beginning, was introduced in \cite{GFThamiltonian}, and an extension of this procedure to the case of a GFT action with four scalar fields was proposed in \cite{Gielen:2020fgi}.
In the latter case, which we summarise in this section, one needs to single out a clock field to construct the Hamiltonian, which breaks the $E(4)$ symmetry between the fields.
In the following we will denote the clock field with $\chi^0$ and the other ``spatial'' fields as $\chi^a$ or $\chi^b$, where $a,b = 1,\,2,\,3$. 
The Hamiltonian associated to  \eqref{GFTLagrange} reads 
\begin{align}
\begin{split}
    H = & \int {\rm d}^3 \chi \sum_J \frac{\mk^{(2)}_J}{2}\left( - \frac{\pi_J^2}{|\mk^{(2)}_J|^2} +  m_J^2 \varphi_J^2 + \sum_b (\pd_b \varphi_J)^2\right)+V(\varphi)\,,
\end{split}
\end{align}
where we introduced the canonical momentum $\pi_J = - \mk_J^{(2)} \pd_0 \varphi_J$ and the shorthand $ m_J^2 = - \frac{\mk_J^{(0)}}{\mk_J^{(2)}}$ . 

Restricting to the free theory with $V(\varphi)=0$ from now on, we then carry out a Fourier decomposition of the above, defining $\omega_{J,k}^2 = m_J^2 + \Vec{k}^2$:
\begin{align}
    H = & \piInt{k} \sum_J \frac{\mkt_J}{2}\left( -\frac{1}{|\mkt_J|^2} \pi_{J, -k}\chiz \pi_{J,k} \chiz + \omega_{J,k}^2\,\varphi_{J, -k}\chiz\varphi_{J,k}\chiz \right)\,.
    \label{eq:hamiltonian}
\end{align}

For the (Heisenberg picture) quantisation we promote the Fourier modes of the group field $\varphi_J$ and its conjugate momentum $\pi_J$ to operators satisfying
\be
[\varphi_{J,k}(\chi^0),\pi_{J',k'}(\chi^0)] = \ii\,\delta_{J J'} (2\pi)^3  \delta(\Vec{k} + \Vec{k'})\,,
\label{eq:commVarphiPi}
\ee
so that these operators evolve in time according to
\begin{align}
    \pd_0 \pi_{J,k} = -\ii\,\left[\pi_{J,k},H\right] = - \mkt_J \omega_{J,k}^2 \varphi_{J,k}\,, \qquad \pd_0 \varphi_{J,k} = - \frac{\pi_{J,k}}{\mkt}\,,
    \label{eq:Qeom}
\end{align}
just as the classical field modes would.

It is then useful to introduce time-dependent creation and annihilation operators $A_{J,k} \chiz,$ $A_{J,k}^\dagger\chiz$, which satisfy  the equal-time commutation relations
\begin{align}
   [A_{J,k}\chiz, A_{J',k'}^\dagger\chiz] = \delta_{J J'} (2\pi)^3 \delta(\Vec{k} - \Vec{k'})\,, 
   \label{eq:commRelTimeDep}
\end{align}
with all other commutators vanishing. These operators are defined from the field operators via
\begin{align}
    \pi_{J,k}\chiz = -{\rm i}\alpha_{J,k}(A_{J,k} - A_{J,-k}^\dagger)\,, \;\; \varphi_{J,k}\chiz = 
    \frac{1}{2\alpha_{J,k}} (A_{J,k} + A_{J,-k}^\dagger)\,; \quad  \alpha_{J,k} = \sqrt{\frac{|\omega_{J,k}| |\mk^{(2)}|}{2}}\,.
     \label{eq:piPhiAaDagger} 
\end{align}

In the following, we will also use time-\emph{independent} (or Schr\"odinger picture) creation and annihilation operators $a_{J,k}, \,  a_{J,k}^\dagger$, defined by  $a_{J,k} = A_{J,k}(0)$ and $a^\dagger_{J,k} = A_{J,k}^\dagger(0)$.

As discussed in detail in \cite{GFThamiltonian,Gielen:2020fgi}, when written in terms of ladder operators the Hamiltonian takes on a different form depending on the sign of $\omega^2_{J,k}$, which in general depends both on the sign of $m_J^2$ and the value of $\vec{k}$. Writing the total Hamiltonian (\ref{eq:hamiltonian}) as $H = \piInt{k} \sum_J H_{J,k}$, for a mode with $\omega_{J,k}^2 < 0$ one finds the Hamiltonian of a harmonic oscillator
\begin{align}
    H_{J,k} = &  -\sgn{\mkt_J} \frac{|\omega_{J,k}|}{2}  \left(a_{J,-k} a_{J,-k}^\dagger + a^\dagger_{J,k} a_{J,k} \right)\,,
    \label{eq:Hosc}
\end{align}
whereas for a mode with $\omega_{J,k}^2 >0$ we obtain a squeezing Hamiltonian
\be
    H_{J,k} = \sgn{\mkt_J} \frac{|\omega_{J,k}|}{2}  \left(a_{J,k} a_{J,-k} + a^\dagger_{J,k} a^\dagger_{J,-k} \right)\,.
    \label{eq:Hsq}
\ee
The modes with squeezing Hamiltonian are of particular interest for cosmology, given that cosmological time evolution can be interpreted as squeezing \cite{Adjei:2017bfm}, or in other words, given that the action of such a Hamiltonian leads to an exponentially growing number of quanta of geometry which one can interpret as an expanding universe. In contrast, a harmonic oscillator Hamiltonian leads to a conserved particle number for the given mode, more akin to a static cosmology.

A particularly natural choice for $\mkt_J$ and $\mkz_J$, which is often considered in the literature (see also our discussion below (\ref{GFTaction})), is obtained from a GFT action whose kinetic term includes a mass term and a Laplace--Beltrami operator on ${\rm SU}(2)^4\times{\mathbb R}^4$. In this case, one may set
\begin{align}
	\mkz_J = \mu - \sum_{i=1}^4  j_i (j_i +1), \quad \mkt_J = \tau\,,
 \label{eq:Kterms}
\end{align}
for some constants $\mu$ and $\tau$, and where the $j_i$ are the irreducible representations appearing in the multi-index $J$. We then have $m_J^2 = \frac{\sum_i j_i (j_i + 1) - \mu }{\tau}$ and, if $\mu>0$, there is a sign change in $m_J^2$ for large $j_i$ values.  Choosing $\tau<0$ would imply that only small $j$ modes have $m_J^2>0$ \cite{Gielen:2016uft}.

For the harmonic oscillator Hamiltonian \eqref{eq:Hosc} the explicit expressions for time-dependent ladder operators are\footnote[-2]{The expressions given in the published version are valid for $\sgn{\mkt_J}<0$, the ones displayed here are general.}  
\begin{align}
    A_{J,k} = a_{J,k} e^{\ii\, \sgn{\mkt_J}|\omega_{J,k}| \chi^0}\,, \quad A_{J,k}^\dagger = a_{J,k}^\dagger e^{-\ii\, \sgn{\mkt_J}|\omega_{J,k}|\chi^0}\,,
    \label{eq:Adynamics}
\end{align} 
whereas for a squeezing Hamiltonian of the form of \eqref{eq:Hsq}, the operator dynamics are solved by 
\begin{align}
    A_{J,k} = &\, a_{J,k} \cosh \left(|\omega _{J,k}| \chi^0\right)-\ii\, \text{sgn}(\mkt) a_{J,-k}^{\dagger } \sinh \left(|\omega _{J,k}| \chi^0\right)\,,\cr
    A_{J,k}^\dagger= &\, a_{J,k}^\dagger \cosh \left(|\omega _{J,k}| \chi^0\right)+\ii\, \text{sgn}(\mkt) a_{J,-k} \sinh \left(|\omega _{J,k}| \chi^0\right)\,.
    \label{eq:aadSolutions}
\end{align}
These expressions can be used to write down the time-dependent expression for all operators constructed out of $A_{J,k} \chiz,$ $A_{J,k}^\dagger\chiz$; within the approximation to the free theory, dynamics can be solved exactly in the Heisenberg picture. We will use this fact in the new results obtained below.

\subsection{Energy-momentum tensor}
\label{sec:GFTTAB}

We now proceed to quantise the components of the GFT energy-momentum tensor $T^{AB}$  \eqref{eq:TGFTdef}, thus introducing a novel set of operators. From hereon we restrict to a single $J$ mode and therefore omit the $J$ label; the single-mode assumption is often used when considering GFT cosmology, as the mode with the largest $|m_J|$ will dominate at late times \cite{Gielen:2016uft}. Here we are not yet interested in a specific physical scenario, but from the form of the Hamiltonian \eqref{eq:hamiltonian} it is apparent that the construction carried out below can be conducted for any of the $J$ modes, and the total operators would be a sum over all modes. It would be straightforward to extend our discussion to multiple modes, which could be interesting, e.g., for phenomenological applications such as \cite{deCesare:2017ynn,Oriti:2021rvm,Calcinari:2022iss}. 

First, we insert the expression for $\pi$ into \eqref{eq:TGFTdef}, where the expressions below depend on the clock as well as the spatial fields $T^{AB} = T^{AB}(\chi^0, \Vec{\chi})$. We find
\begin{align}
\begin{split}
    T^{00} =&\,\frac{\pi^2}{2\mk^{(2)}} - \frac{\mk^{(2)}}{2}\left(m^2\varphi^2  +\sum_b (\pd_b\varphi)^2\right) \,,\\
    T^{0 b} = &  -  \pi \pd_b \varphi\,, \qquad  T^{a \neq b} =  \mk^{(2)}\pd_a \varphi \pd_b \varphi\,,\\
    T^{aa} = &\,  - \frac{\pi^2}{2\mk^{(2)}}-\frac{\mk^{(2)}}{2}\left(m^2\varphi^2-(\pd_a \varphi)^2 + \sum_{b\neq a}(\pd_b \varphi)^2 \right)\quad({\rm no\;sum\;over\;}a)\, . \label{eq:GFTT_cl}
\end{split}
\end{align}
The Fourier transforms $T^{AB}_k = T^{AB}_k(\chi^0)$ of the above functions, where $\varphi_k = \varphi_k(\chi^0)$ and $\pi_k = \pi_k(\chi^0)$ denote the Fourier transforms of $\varphi(\chi^0, \Vec{\chi})$ and $\pi(\chi^0, \Vec{\chi})$, respectively, are given by the following convolutions:
\begin{align}
\begin{split}
   T_k^{00} = & \,\frac{1}{2}\piInt{\gamma} \Bigl[\frac{\pi_\gamma \pi_{k-\gamma}}{\mkt} -\mkt \left(m^2 - \vec\gamma\cdot(\vec{k} - \vec\gamma)\right)\varphi_\gamma \varphi_{k-\gamma}\Bigr]\,,\\
   T_k^{0b} = & -{\rm i}\piInt{\gamma}  \gamma_b \pi_{k-\gamma}\varphi_{\gamma}\,, \qquad 
   T_k^{a\neq b} =  -\piInt{\gamma}  \mkt \gamma_a (k_b - \gamma_b)\varphi_\gamma \varphi_{k - \gamma}\,,\\  
   T_k^{a a} = & \,\frac{1}{2}\piInt{\gamma}  
   \Bigl[ \mkt \left(- \gamma_a(k_a - \gamma_a) + \sum_{b\neq a}\gamma_b(k_b - \gamma_b)  -m^2 \right)\varphi_\gamma \varphi_{k - \gamma} - \frac{\pi_\gamma \pi_{k - \gamma}}{\mkt}\Bigr]
  \,.
   \label{eq:TfourierCl}
\end{split}
\end{align}
Initially, we will think of the corresponding operators, which we denote $\T^{AB}$, as defined simply by replacing the Fourier modes $\varphi_k$ and $\pi_k$ by operators, with $\pi_{k-\gamma}$ appearing to the left of $\varphi_\gamma$ in $\T_k^{0b}$. (We discuss normal ordering below.)

The classical energy-momentum tensor satisfies $\pd_0 T_k^{0B} + {\rm i} \sum_a  k_a T_k^{aB} = 0$. One would expect this conservation law to also hold at the level of operators: any alterations could only arise from commutators \eqref{eq:commVarphiPi} that appear from necessary operator reordering. However, such terms are always proportional to $\delta(\vec{k})$ and time-independent, so that we obtain $\pd_0 \T^{0B}_k + \ii \sum_a k_a \T^{aB}_k  = \pd_0 \xi^{(0)} \delta(\vec{k}) + \ii \sum_a \xi^{(a)}k_a \delta(\vec{k}) = 0$, where $\xi^{(A)}$ are independent of $\chi^0$. Explicitly, we find
\begin{align}
\begin{split}
     \pd_0 \T_k^{00} = & \piInt{\gamma}\frac{1}{2}\Bigg[
     \left(\frac{\pd_0 \pi_\gamma}{\mkt} +\left(m^2 - \Vec{\gamma}\cdot(\Vec{k} - \Vec{\gamma})\right) \varphi_\gamma\right) \pi_{k-\gamma} + \pi_{k-\gamma}\left(\frac{\pd_0 \pi_{\gamma}}{\mkt} + \left(m^2 -  \Vec{\gamma}\cdot(\Vec{k} - \Vec{\gamma})\right)\varphi_{\gamma}\right)
     \Bigg]\,,\\
   \ii \sum_a k_a \T_k^{0a} = & \piInt{\gamma}\frac{1}{2}\Vec{k}\cdot\Vec{\gamma} \left(\pi_{k-\gamma}\varphi_\gamma +\varphi_\gamma\pi_{k-\gamma} + [\pi_{k-\gamma}, \varphi_\gamma] \right)\,,
\label{eq:consLzeroComponents}
\end{split}
\end{align}
where we made use of $\piInt{\gamma} f(\vec\gamma) g(\vec{k}-\vec\gamma) = \piInt{\gamma} f(\vec{k}-\vec{\gamma}) g(\vec{\gamma})$.

Combining the above terms, we then obtain
\begin{align}
\begin{split}
     \pd_0 \T_k^{00} + & \ii \sum_a k_a \T_k^{0a}  \\
    = & \piInt{\gamma}\frac{1}{2} \Bigg[\left(  \frac{\pd_0 \pi_\gamma}{\mkt} + \omega_\gamma^2 \varphi_\gamma \right)\pi_{k-\gamma} + \pi_{k-\gamma}\left(  \frac{\pd_0 \pi_\gamma}{\mkt} + \omega_\gamma^2 \varphi_\gamma \right)+ \vec{k}\cdot\vec\gamma [\pi_{k-\gamma}, \varphi_\gamma] \Bigg]\\
    =& \,0
\end{split}
\end{align}
from \eqref{eq:Qeom} and $[\pi_{k-\gamma}, \varphi_\gamma] \propto \delta(\vec{k})$. As anticipated, the extra term coming from operator ordering does not contribute, as it is proportional to $\Vec{k}\delta(\Vec{k}) = 0$.

In the case of $\pd_0 \T_k^{0b} + {\rm i} \sum_a  k_a \T_k^{ab} = 0$ we have 
\begin{align}
    \pd_0 \T_k^{0b} = & \,\ii \piInt{\gamma} \gamma_b \left(\mkt  \omega_{k-\gamma}^2 \varphi_{k-\gamma}\varphi_\gamma + \frac{\pi_{k-\gamma }\pi_\gamma}{\mkt} \right)\,,\nonumber\\
     {\rm i} \sum_a  k_a \T_k^{ab} = & \,\ii \piInt{\gamma} \Bigg[
    \left( - (k_b - \gamma_b)\Vec{k}\cdot\Vec{\gamma} + \frac{k_b}{2}(\Vec{k} - \Vec{\gamma})\cdot\Vec{\gamma}-\frac{m^2}{2}k_b\right) \mkt\varphi_{\gamma}\varphi_{k-\gamma}- \frac{k_b}{2}\frac{\pi_\gamma \pi_{k-\gamma}}{\mkt}
     \Bigg]\,.
\end{align}
In this case there are no mixed $\pi\varphi$ terms, so there can be no nontrivial commutators and the conservation law should follow as for the classical fields. To see this, in the expression for $\pd_0 \T_k^{0b}$ change integration variables $\vec\gamma\rightarrow\vec{k}-\vec{\gamma}$, add to the original expression and divide by 2 to obtain
\begin{align}
\pd_0 \T_k^{0b} =&\, \frac{\ii}{2} \piInt{\gamma} \left[\left(\gamma_b \omega_{k-\gamma}^2 + (k_b-\gamma_b)\omega_\gamma^2\right)\mkt\varphi_{k-\gamma}\varphi_\gamma + k_b \frac{\pi_{k-\gamma }\pi_\gamma}{\mkt}\right]
\cr = & \, \frac{\ii}{2} \piInt{\gamma} \left[\left(\gamma_b (\vec{k}-\vec{\gamma})^2 + (k_b-\gamma_b)\vec\gamma^2+k_b m^2\right)\mkt\varphi_{k-\gamma}\varphi_\gamma + k_b \frac{\pi_{k-\gamma }\pi_\gamma}{\mkt}\right]\,.
\end{align}
We then obtain
\be
\pd_0 \T_k^{0b}+{\rm i} \sum_a  k_a \T_k^{ab} = \frac{\ii}{2} \piInt{\gamma} \left[\left(\gamma_b \vec{k}^2 -k_b(\vec{k}\cdot\vec\gamma)\right)\mkt\varphi_{k-\gamma}\varphi_\gamma \right] = 0\, ,
\ee
since the integral can be transformed into minus itself under a variable change $\vec\gamma\rightarrow\vec{k}-\vec{\gamma}$.

We have hence shown explicitly that the operators representing the Fourier modes of the GFT energy-momentum tensor for a single mode satisfy the conservation law $\pd_0 \T_k^{0B} + {\rm i} \sum_a  k_a \T_k^{aB} = 0$. We should stress again that this is true already for the operators, and would in particular hold for the corresponding expectation values in {\em any} state.

Here we have defined the energy-momentum tensor without applying a normal ordering prescription. One might then be worried that there are divergences in this definition of $\T^{AB}_k$, and wonder whether the same conservation law applies to a normal-ordered definition $:\T^{AB}_k\!:$. Normal ordering can be implemented in terms of the time-dependent operators $A_{k} \chiz$ and $A_{k}^\dagger\chiz$, i.e., moving all $A_{k} \chiz$ to the right of any $A_{k}^\dagger\chiz$, if one wants the procedure to be equivalent to standard normal ordering in the Schr\"odinger picture. However, this is not enough to cancel divergences coming from the unstable squeezed modes with Hamiltonian (\ref{eq:Hsq}). In \cite{Gielen:2020fgi}, normal ordering for time-dependent operators was followed by a regularisation in which vacuum expectation values were subtracted. Both steps can be implemented as one by imposing normal ordering at the level of the $a_k$ and $a_k^\dagger$ operators, leading to the definition
\begin{align}
\begin{split}
    :\T^{00}_k\!: \; = & \piInt{\gamma}  \frac{\text{sgn}(\mkt)}{4 \sqrt{|\omega_\gamma||\omega_{k-\gamma}|}}\Bigg[ 2 \beta^+_{k, \, \gamma}  :A^{\dagger }_{-\gamma }A_{k-\gamma }: 
    + \beta^-_{k,\,\gamma} \left( :A^{\dagger }_{-\gamma }A^{\dagger }_{\gamma -k}: + :A_{\gamma }A_{k-\gamma }:
   \right)
   \Bigg]\,,\\
   :\T^{0b}_k\!: \; = & \piInt{\gamma} \frac{1}{2}\sqrt{\frac{|\omega_{k-\gamma}|}{|\omega_\gamma|}} \gamma_b
   \left(  :A^{\dagger }_{\gamma
   -k}A_{\gamma }: - :A_{-\gamma }^{\dagger }A_{k-\gamma }:
   - :A_{k-\gamma }A_{\gamma }:
   + :A^{\dagger }_{\gamma -k}A^{\dagger }_{-\gamma }:
   \right)\,,\\
   :\T^{a\neq b}_k\!: \; = & \piInt{\gamma}  \frac{\text{sgn}(\mkt)}{2 \sqrt{|\omega_\gamma||\omega_{k-\gamma}|}}\gamma _a \left(\gamma _b-k_b\right)
   \left( :A^{\dagger }_{-\gamma }A_{k-\gamma }: + :A_{\gamma -k}^{\dagger }A_{\gamma }: + :A_{-\gamma }^{\dagger }A_{\gamma -k}^{\dagger }: + :A_{\gamma }A_{k-\gamma }:\right)\,,\\
   :\T^{aa}_k\!: \; = &  \piInt{\gamma} \frac{\text{sgn}(\mkt)}{4 \sqrt{|\omega_\gamma||\omega_{k-\gamma}|}} \Bigg[ 2(\beta_{k ,\gamma}^- - 2 \gamma_a(k_a - \gamma_a)) :A^{\dagger }_{-\gamma }A_{k-\gamma }: \cr
   & \qquad + (\beta_{k ,\gamma}^+ - 2 \gamma_a(k_a - \gamma_a))\left(:A^{\dagger }_{-\gamma }A^{\dagger }_{\gamma -k}: + :A_{\gamma }A_{k-\gamma }: \right)  \Bigg]\,,\
\end{split}
\label{eq:TABoperatorsNormal}
\end{align}
where we defined $\beta_{k ,\gamma}^{\pm} =  -m^2 + \Vec{\gamma}\cdot(\Vec{k}-\Vec{\gamma})  \pm |\omega_\gamma||\omega_{k-\gamma}|$.

Using (\ref{eq:Adynamics}) and (\ref{eq:aadSolutions}), one can now write the Fourier modes of the energy-momentum tensor in terms of time-dependent functions of ladder operators $a_k$ and $a_k^\dagger$, and implement normal ordering. 
Notice that in general only four independent combinations of ladder operators are needed to define all components of $\T^{AB}$.
However, given that in general there are two different types of modes (squeezed, $\omega_k^2>0$, and oscillating, $\omega_k^2<0$, ones), these explicit expressions  depend on what types of mode contribute, and hence on the value of $m^2$. If $m^2<0$ and for $\vec\gamma^2<|m^2|$, $A_{\gamma}$ and $A_{\gamma}^\dagger$ operators have the dynamics of oscillating modes \eqref{eq:Adynamics}; for all other cases they follow the dynamics of squeezing modes $\eqref{eq:aadSolutions}$. Hence, in  the operator products appearing in the expressions for $T^{AB}$ both operators can be of squeezing type, one can be  of squeezing and one of oscillating type, or both can be oscillating modes. The mixed case occurs for all $\vec{k}\neq 0$ if $m^2<0$, but in the following we will only be interested in $\vec{k}=0$ expressions, and therefore only provide the expressions for operator pairs of the same type; it is straightforward to derive the mixed cases from \eqref{eq:Adynamics} and \eqref{eq:aadSolutions}.

For pairs of operators associated to two oscillating modes ($\omega_\gamma^2 < 0$ and $\omega_{k-\gamma}^2 <0$) we obtain the simple form\footnote[-3]{Again (see footnote -2) the expressions given here generalise the ones given in the published version.}
\begin{align}
\begin{split}
     \no{A_{-\gamma}^\dagger A_{k-\gamma}} \  = & \; a^\dagger_{-\gamma} a_{k-\gamma}\,,\\ 
    \no{A_{\gamma-k}^\dagger A_{\gamma}} \ = &   \; a^\dagger_{\gamma-k}a_\gamma\,, 
\end{split}
\begin{split}
    \no{A_{-\gamma}^\dagger A_{\gamma-k}^\dagger}  \ = & \; a_{-\gamma}^\dagger a_{\gamma -k}^\dagger e^{-\ii\, \sgn{\mkt_J} (|\omega_{-\gamma}| + |\omega_{\gamma - k}|)\chi^0}\,, \\  
    \no{A_{\gamma} A_{k-\gamma}} \ = & \; a_\gamma a_{k-\gamma} e^{\ii \,\sgn{\mkt_J}(|\omega_{\gamma}| + |\omega_{ k- \gamma }|)\chi^0}\,.
\end{split}
    \label{eq:aadSol_osc}
\end{align}

Normal-ordered expressions for pairs of $A $ and $A^\dagger$ operators associated to two squeezed modes ($\omega_\gamma^2 > 0$ and $\omega_{k-\gamma}^2 >0$) are given by \footnote[-4]{Here we have introduced absolute value signs compared to the published version, which assumes $\omega_\gamma>0$, $\omega_{k-\gamma}>0$.}
\begin{align}
\begin{split}
    \no{A_{-\gamma}^\dagger A_{k-\gamma}} \  = & \;a^\dagger_{\gamma
   -k} a_{\gamma } \sinh \left( |\omega _{-\gamma }| \chi^0 \right) \sinh \left(|\omega _{k-\gamma }| \chi^0\right)+ a^\dagger_{-\gamma }a_{k-\gamma } \cosh \left(|\omega _{-\gamma }| \chi^0\right) \cosh \left(|\omega
   _{k-\gamma }| \chi^0\right)\\
   & + \ii \,\sgn{\mkt} \left(a_{\gamma }a_{k-\gamma } \sinh \left( |\omega _{-\gamma }| \chi^0\right) \cosh \left( |\omega _{k-\gamma }| \chi^0\right)-
   a^\dagger_{-\gamma } a^\dagger_{\gamma -k} \cosh \left(|\omega _{-\gamma }| \chi^0 \right) \sinh \left(|\omega _{k-\gamma }| \chi^0\right)\right)\\
   = &\;  A_{-\gamma}^\dagger A_{k-\gamma} - \sinh(\omega_\gamma \chi^0)^2 (2\pi)^3 \delta(\vec{k})\,,\\
    \no{A_{\gamma-k}^\dagger A_{\gamma}} \ = &  \; a^\dagger_{-\gamma } a_{k-\gamma } \sinh \left(|\omega _{\gamma }|\chi^0\right) \sinh \left(|\omega _{\gamma -k}| \chi^0\right)+  a^\dagger_{\gamma -k} a_{\gamma } \cosh \left(|\omega _{\gamma   }| \chi^0\right) \cosh \left(|\omega _{\gamma -k}| \chi^0\right)\\
     & + \ii\, \sgn{\mkt} \left( -a^\dagger_{\gamma -k} a^\dagger_{-\gamma }\sinh \left(|\omega _{\gamma }| \chi^0\right) \cosh \left(|\omega
   _{\gamma -k}| \chi^0\right) + a_{k-\gamma }a_{\gamma }\cosh \left(|\omega _{\gamma }| \chi^0\right) \sinh
   \left(|\omega _{\gamma -k}| \chi^0\right) \right)\\
    = &\; A_{-\gamma}^\dagger A_{k-\gamma}  - \sinh(\omega_\gamma \chi^0)^2 (2\pi)^3 \delta(\vec{k})\,,\\
    \no{A_{-\gamma}^\dagger A_{\gamma-k}^\dagger} \ = & \; a^\dagger_{-\gamma } a^\dagger_{\gamma -k} \cosh \left(|\omega _{-\gamma }| \chi^0\right) \cosh \left(|\omega _{\gamma -k}| \chi^0\right) -   a_{\gamma }a_{k-\gamma } \sinh \left(|\omega_{-\gamma }|\chi^0\right) \sinh \left(|\omega _{\gamma -k}| \chi^0\right) \\,
   & + \ii\, \sgn{\mkt} \left( a^\dagger_{-\gamma }a_{k-\gamma } \cosh \left(|\omega _{-\gamma }|\chi^0\right) \sinh \left(|\omega
   _{\gamma -k}|\chi^0\right) +    a^\dagger_{\gamma -k} a_{\gamma } \sinh \left(|\omega _{-\gamma }|\chi^0\right) \cosh \left(|\omega _{\gamma -k}|\chi^0\right) \right)\\
   = & \; A_{-\gamma}^\dagger A_{\gamma-k}^\dagger  - \ii \,\sgn{\mkt} \sinh(|\omega_\gamma| \chi^0)\cosh(|\omega_\gamma| \chi^0) (2\pi)^3 \delta(\Vec{k})\,,\\
    \no{A_{\gamma} A_{k-\gamma}} \ = & \; - a^\dagger_{-\gamma }a^\dagger_{\gamma -k} \sinh \left(|\omega _{\gamma }|\chi^0\right) \sinh \left(|\omega _{k-\gamma }|\chi^0\right) + a_{\gamma }a_{k-\gamma } \cosh \left(|\omega_{\gamma }| \chi^0\right)  \cosh \left(|\omega _{k-\gamma }|\chi^0\right)\\
   & - \ii \,\sgn{\mkt}\left(  a^\dagger_{-\gamma } a_{k-\gamma }  \sinh \left(|\omega _{\gamma }|\chi^0\right) \cosh \left(|\omega_{k-\gamma }|\chi^0\right)+ a^\dagger_{\gamma -k} a_{\gamma
   }\cosh \left(|\omega _{\gamma }|\right) \sinh \left(|\omega _{k-\gamma }|\chi^0\right)\right)\\
   = &\; A_{\gamma} A_{k-\gamma}  + \ii \,\sgn{\mkt} \sinh(|\omega_\gamma| \chi^0)\cosh(|\omega_\gamma| \chi^0) (2\pi)^3 \delta(\Vec{k})\,.
\end{split}
\end{align}

The normal ordering procedure only affects the form of the $\T^{00}$ and $\T^{aa}$ operators (the terms arising from operator reordering vanish for the other components), and the difference depends on the sign of $\omega_\gamma^2$ and hence on the type of mode:
\begin{align}
    :\T^{00}_k\!:\; = &\,  \T^{00}_k - \delta(\vec{k}) \frac{\sgn{\mkt}}{4} \int \dd^3\gamma\, |\omega_\gamma| (1 - \sgn{\omega_\gamma^2})\,,\\
    :\T^{aa}_k\!:\; = &\, \T^{aa}_k + \delta(\vec{k}) \frac{\sgn{\mkt}}{4} \int \dd^3\gamma \left( |\omega_\gamma| (1 + \sgn{\omega_\gamma^2}) - \frac{2 \gamma_a^2}{|\omega_\gamma|} \right)\cr
    & \quad + \delta(\vec{k})\, \sgn{\mkt}\int \dd^3\gamma \;\Theta(\omega_\gamma^2)\left( |\omega_\gamma|  - \frac{\gamma_a^2}{|\omega_\gamma|} \right)\sinh^2(\omega_\gamma\,\chi^0)\,.
\end{align}
Given that $\omega_\gamma^2 \sim \vec\gamma^2$ at large $|\vec\gamma|$, the integrals appearing as the difference between the normal-ordered $:\T^{aa}_k\!:$ and the previous definition (\ref{eq:TfourierCl}) are divergent and require regularisation, e.g., by a cutoff. The time-dependent integral in the last line (which involves the Heaviside function $\Theta(\omega_\gamma^2)$ since it only comes from squeezed modes), is particularly badly divergent with an integrand growing exponentially at large $|\vec\gamma|$ for any $\chi^0\neq 0$. 
The integral appearing in $:\T^{00}_k \! :$ is  only relevant for $m^2<0$ and finite, given that there is always a finite $R$ such that $\omega_\gamma^2>0$ for $|\vec\gamma|>R$, and modes with $\omega_\gamma^2>0$ have a vanishing ``vacuum energy'' already before normal ordering. Even this finite integral is however still multiplied by a delta distribution.
If $m^2>0$, this integral vanishes entirely. 
The additional term appearing in the normal ordered $:\T^{00}_k\!:$ is time-independent, whereas the terms in $:\T^{aa}_k\!:$ are multiplied by $\delta(\vec{k})$. Hence, none of these terms contribute to the conservation law and $\pd_0 :\T_k^{0B}: + \,{\rm i} \sum_a  k_a :\T_k^{aB}:\; = 0$ also for the normal-ordered definition, as expected.

A particular case of the conservation law applies to the zero modes of $\T^{0B}$, which satisfy $\pd_0 \T_0^{00}=\pd_0 \T_0^{0a}=0$. These are the usual global conserved quantities corresponding to the total energy and total momentum respectively, which were discussed already in \cite{Gielen_2020}. We will encounter them again in our explicit example below.

This concludes the discussion of the GFT energy-momentum tensor $\T^{AB}$. We considered its conservation law, which holds irrespective of operator ordering, and obtained explicit forms in terms of ladder operators whose dynamics depend on the type of mode we consider: using (\ref{eq:Adynamics}) and (\ref{eq:aadSolutions}) one can write down the explicit time-dependent form of \eqref{eq:TABoperatorsNormal}.  Through our earlier identification of the expectation value of $\T^{AB}$ with the classical current $j^{AB}$, $\langle \T^{AB} \rangle = j^{AB}$, 
and hence the spacetime metric via (\ref{inversemetric}), 
these solutions can be used to define an effective metric from any GFT state that is sufficiently semiclassical (as we will discuss more concretely in section \ref{sec:effectiveFLRW}).
In what follows we examine the implications of this generic construction for a specific example.

\section{Simple cosmology example}
\label{sec:cosmoEx}

As a first consistency check, we apply the above construction to a scenario widely studied in existing GFT literature: the flat Friedmann--Lema\^{i}tre--Robertson--Walker (FLRW) spacetime. 
We begin by recalling the classical scenario, which differs from standard cosmological models due to the four massless scalar fields required to construct the relational coordinate system (see also \cite{Jercher:2023nxa} for a discussion of cosmology with massless scalar fields used as coordinates).

We then take expectation values of the operators \eqref{eq:TABoperatorsNormal} in a highly peaked Gaussian state, which, in addition to fulfilling the semiclassicality requirement, will be shown to represent an effective FLRW geometry.
These expectation values can then be compared to the components of the classical current \eqref{jcurrent} and their dynamics.

\subsection{Classical dynamics}

We first recall the line element of a flat FLRW universe
\begin{align}
    \dd s^2 = -N^2(t) \dd t^2 + a^2(t) \delta_{ij}\dd x^i \dd x^j,
    \label{eq:FLRW}
\end{align}
where $N(t)$ is the lapse function and $a(t)$ the scale factor. 
As described in section~\ref{sec:current}, we construct the quantum theory in a relational coordinate system characterised by  $\pd_\mu \chi^A = \delta_\mu^A$ and any identification of classical quantities with operator expectation values holds only for this choice of coordinates. The units of the lapse and scale factor are $[a] = [N] = L^2$ and $[\pd_\mu \chi^A] = L^0$. The lapse for this choice of coordinate system can be obtained from the definition of the canonical momentum $\pi_0$ conjugate to $\chi^0$, namely $ \pi_0 = \frac{a^3}{N} (\pd_t\chi^0) \Rightarrow N = \frac{a^3}{\pi_0}$. (The momenta of the spatial fields give the shift vector as $N^a = - \frac{\pi_a}{\pi_0}$, which vanishes for \eqref{eq:FLRW}, so these momenta must vanish too.) The Klein--Gordon equation for $\chi^0$ with FLRW symmetry is equivalent to the statement $\partial_t\pi_0=0$.

In a flat FLRW universe, the conserved currents \eqref{jcurrent} thus take the form\footnote[-5]{Recall (see footnote -1) that in our gauge $\pi_0 > 0$, so the signature of $j^{AB}$ is always $(+---)$ as required.}
\be
j^{AB} = \begin{pmatrix} \pi_0 & 0 \cr 0 & -\frac{a^4}{\pi_0}\delta^{ab}
\end{pmatrix}\,.
\label{eq:jAB_bg}
\ee
Notice that all components have a fixed sign, determined by the Lorentzian signature implemented in (\ref{eq:FLRW}). For Euclidean signature, all entries would be positive.

The classical energy-momentum tensor of the four scalar fields, which we will denote as $\tensor{\Tmat}{^\mu_\nu}$ to avoid confusion with the GFT energy-momentum tensor $T^{AB}$, differs from that of the single matter field case. It contains contributions $\propto a^{-2}$ from the spatial fields, as their derivatives are non-vanishing at background level:
\begin{align}
    -\tensor{\Tmat}{^0_0} =\rho  = \frac{\pi_0^2}{2 a^6} + \frac{3}{2 a^2}\,, \qquad  \tensor{\Tmat}{^a_a} = P = \frac{\pi_0^2}{2 a^6} - \frac{1}{2 a^2}\,.
\end{align}
The gradient energy coming from the spatial coordinate fields appears as an additional term that would be equivalent to negative spatial curvature.

The resulting first and second Friedmann equations read (with $\kappa=8\pi G$, and $'$ denoting derivatives with respect to $\chi^0$)
\begin{align}
    H^2 & = \left(\frac{a'}{a}\right)^2 = \frac{\kappa}{6} \left( 1 + 3 \frac{a^4}{\pi_0^2} \right), \qquad \frac{a''}{a} = \frac{\kappa}{6} \left( 1 + 9 \frac{a^4}{\pi_0^2} \right),
\label{eq:clFried}
\end{align}
where, again, the terms proportional to $\frac{a^4}{\pi_0^2}$ would not appear in the case of a single (clock) scalar field. 
This also implies that we no longer have an equation of state parameter $w = \frac{P}{\rho}$ exactly equal to one, but instead $w  < 1$. 
The contribution of the spatial fields to the energy density and pressure becomes negligible in the limit where $\frac{\pi_0^2}{a^4} \gg 1$, effectively recovering the standard cosmological background scenario with a massless scalar field. 
Similarly, for the sound speed we find $c_s^2 = \frac{P'}{\rho'} = 
\frac{3 \pi_0^2 - a^4}{3\pi_0 + 3 a^4}$, and thus $c_s^2 \approx 1$ if we again assume $\frac{\pi_0^2}{a^4} \gg 1$.
This limit can be achieved for sufficiently early times, depending on the value of $\pi_0$, but at late times the gradient energy will dominate.

The scenario we have described here is rather unusual from the viewpoint of conventional homogeneous cosmology, where one would require the fields $\chi^A$ to be spatially homogeneous, such that there can be no contributions from gradient energy. Of course, such a symmetry requirement is incompatible with the condition that the $\chi^a$ can be spatial coordinates, $\partial_i\chi^a=\delta_i^a$. Here we adopted the view that only observables such as the energy-momentum tensor need to be spatially homogeneous, and given that the field values themselves never enter any observables for free massless scalar fields, this is classically consistent with nonvanishing gradient energy. 
Similar ideas appear in the context of solid inflation \cite{Endlich:2012pz} where an analogous gradient energy contribution is present in the Friedmann equation.
In the context of GFT, these assumptions may indeed be mutually incompatible, as we discuss shortly.

\subsection{Effective FLRW metric from GFT energy-momentum tensor}
\label{sec:effectiveFLRW}

In order to extract an effective metric, we need to determine a  state which reflects the physical scenario we are interested in, as well as being sufficiently semiclassical.
As shown in \cite{Gielen_2020} for models with a single massless scalar field, Fock coherent states form a suitable class of semiclassical states, as the relative uncertainty in the volume remains small throughout the evolution (see also \cite{Calcinari:2023sax} for a more in-depth analysis of a broader class of semiclassical GFT states). It is then reasonable to assume that the same holds for expectation values of the $\T^{AB}$ operators.
We therefore use a coherent state $\ket{\sigma}$ which is an eigenstate of the (time-independent) annihilation operator $ a_{J,k}\ket{\sigma} = \sigma_J(\Vec{k}) \ket{\sigma}$: 
\begin{align}
   \ket{\sigma} = e^{- ||\sigma||^2/2} \exp\left(\sum_J \int \frac{{\rm d}^3k}{(2\pi)^3} \sigma_J(\Vec{k}) a^\dagger_{J,k}\right)|0\rangle\,,
    \label{eq:sigmaGauss}
\end{align}
where $|0\rangle$ is the GFT Fock vacuum and $||\sigma||^2 = \sum_J \piInt{k} |\sigma_J(\Vec{k})|^2$.

We are interested in flat FLRW cosmology, where all quantities, in particular the components of the classical current $j^{AB}$, are homogeneous. A cosmological quantum state should reflect this homogeneity.
We thus choose a Gaussian as a sharply peaked mean field,
\begin{align}
  \sigma_J(\Vec{k}) = \delta_{J,J_0}\frac{\ma + \ii \mb}{c_\sigma} \ e^{-\frac{(\Vec{k}-\Vec{k}_0)^2}{2 s^2}} \, ,
\end{align}
where $\ma, \, \mb \in \mathbb{R}$, $s $ determines the peakedness of the state and $\Vec{k}_0$ is the initially dominantly excited Fourier mode.  
Similar to the single mode restriction made in sec.\,\ref{sec:GFTTAB}, in what follows we assume only one Peter--Weyl mode with $J=J_0$ is excited. This is a standard assumption in the literature \cite{Gielen_2020,Marchetti:2021gcv}, motivated by the fact that in an expanding universe a single mode will dominate at late times and thereby determine the late-time limit \cite{Gielen:2016uft}. 
(In general, if multiple modes are included, the initial condition parameters $\ma, \ \mb$ and $s$ could of course  be $J$ dependent.) We have fixed the normalisation factor $c_\sigma = \left(\frac{s}{2 \sqrt{\pi}}\right)^{3/2}$ for convenience regarding later calculations.
As the $\Vec{k}=0$ mode corresponds to the homogeneous mode, a strictly homogeneous state would correspond to an infinitely peaked state around $\Vec{k}_0 = 0$, i.e., the limit $s \to 0$. To avoid any divergences, we choose a state with small, but finite $s$. This introduces a conceptual discrepancy to the classical case: in the quantum theory it is not possible to excite solely the homogeneous background, but inhomogeneous modes will always be excited to some degree, which clearly differs from the standard distinction between background and perturbations in cosmology.
In the following we set $\Vec{k}_0 = 0$ and consider the dynamics of the homogeneous, $\Vec{k}=0$ mode; all other modes should be identified with perturbations of the homogeneous background (which we will study in detail in an upcoming article). 

The single mode $J_0$ we consider can either have $m^2_{J_0} = m^2 >0$ or $m^2<0$ (we ignore the fine-tuned special case $m^2=0$ which needs to be analysed separately \cite{Gielen:2023han}). Since the mean field is sharply peaked at $\vec{k}=0$, the expectation value of the energy-momentum tensor will be determined by (low $|\vec{k}|$) squeezed modes for $m^2>0$ and  by  (low $|\vec{k}|$) oscillating modes for $m^2<0$.

 For $m^2>0$ we obtain the following expectation values:
\begin{align}
\begin{split}
    \ev{\T^{00}_0} = & \piInt{\gamma} \sgn{\mkt}  |\omega_\gamma|(\mb^2 - \ma^2)  \frac{e^{-\gamma^2/s^2}}{c_\sigma^2} 
    \approx \sgn{\mkt} |m| (\mb^2 - \ma^2)\,, \\ 
    \ev{\T^{0b}_0} = & \; 0\, ,  \qquad     \ev{\T^{a\neq b}_0} = 0\,,\\
    \ev{\T^{aa}_0} = & \piInt{\gamma} \sgn{\mkt} \frac{e^{-\gamma^2/s^2}}{c_\sigma^2} \\
    & \quad \times \left( \left(- |\omega_\gamma| + \frac{\gamma_a^2}{|\omega_\gamma|}\right) \left( (\ma^2 + \mb^2)\cosh(2 \omega_\gamma \chi^0) - 2\, \sgn{\mkt} \ma \mb  \sinh(2 \omega_\gamma \chi^0) \right) + \frac{\gamma^2_a}{|\omega_\gamma|} (\ma^2 - \mb^2)\right)  \\
    \approx &   - \sgn{\mkt}   |m| \left( (\ma^2 + \mb^2)\cosh(2 |m| \chi^0) - 2\, \sgn{\mkt} \ma \mb  \sinh(2 |m| \chi^0) \right)\,. 
\label{eq:TAB_bg}
\end{split}
\end{align}
To simplify the integrals appearing in $\ev{\T^{00}_0}$ and $\ev{\T^{aa}_0}$\footnote{The integral for $\ev{\T^{00}_0}$ can be carried out analytically to give a Tricomi confluent hypergeometric function, but as this does not add value to the results we present, we omit this result.},  we used the saddle-point approximation 
\begin{align}
    \int \dd^3 x \;e^{-\frac{(\Vec{x}-\Vec{\mu} )^2}{s^2}} g(\Vec{x}) \approx  g(\Vec{\mu}) \int \dd^3 x \;e^{-\frac{(\Vec{x}-\Vec{\mu} )^2}{ s^2}} = g(\Vec{\mu}) (\sqrt{\pi} s)^3,
    \label{eq:saddlePtApproxLO}
\end{align}
which holds for sharply peaked Gaussians such that $g(\vec{x})$ can be considered approximately constant in the region $|\vec{x}-\vec{\mu}|\le s$. In effect, this approximation corresponds to the idealised limit of $s\rightarrow 0$ given that we are ignoring all finite $s$ contributions; we discuss the effects of next-to-leading-order contributions in appendix \ref{app:saddlePointNLO}, demonstrating that they can rightfully be neglected. The approximation will break down at late times for $\ev{\T^{aa}_0}$, as in the similar discussion of \cite{Gielen:2020fgi}.
The integrals for $ \ev{\T^{0b}_0}$ and $ \ev{\T^{a\neq b}_0}$ vanish due to antisymmetry. This result for $\ev{\T^{0b}_0}$ can be read either as the vanishing of some off-diagonal metric components or of the canonical momenta conjugate to $\chi^b$ (see the discussion above \eqref{eq:jAB_bg}).

For $m^2 <0$, the integral over $\vec\gamma$ will contain both squeezed and oscillating modes, but for a very sharply peaked mean field, only the region near $\vec\gamma=0$, which consists of oscillating modes only, can contribute. We can hence write, using again the saddle-point approximation, \footnote[-6]{Again the expressions given here generalise the ones given in the published version, which apply for $\sgn{\mkt}<0$.}
\begin{align}
\begin{split}
    \ev{\T^{00}_0} \approx & \piInt{\gamma} \sgn{\mkt} \frac{e^{-\gamma^2/s^2}}{c_\sigma^2} |\omega_\gamma| (\ma^2 + \mb^2)    %
    \approx \sgn{\mkt} |m| (\ma^2 + \mb^2)\,, \\ 
    \ev{\T^{0b}_0} = &\; 0\,,  \qquad     \ev{\T^{a\neq b}_0} = 0\,,\\
    \ev{\T^{aa}_0} \approx & \piInt{\gamma} \frac{\sgn{\mkt}}{|\omega_\gamma|}\frac{e^{-\gamma^2/s^2}}{c_\sigma^2} \\
    & \quad \times \left( \gamma_a^2 (\ma^2 + \mb^2) + (|\omega_\gamma|^2 + \gamma_a^2) \left( (\ma^2 - \mb^2) \cos(2 |\omega_\gamma| \chi^0) - 2 \,\sgn{\mkt}\ma \mb \sin(2|\omega_\gamma| \chi^0)\right)\right)\\
    \approx & \;
    \sgn{\mkt} |m|  \left( (\ma^2 - \mb^2) \cos(2 |m| \chi^0) - 2 \,\sgn{\mkt}\ma \mb \sin(2 |m| \chi^0)\right)\,.
\label{eq:TAB_bg_osc}
\end{split}
\end{align}

Before we can proceed to use the identification $j^{AB} = \ev{\T^{AB}}$ to extract an effective metric from the expressions \eqref{eq:jAB_bg} and \eqref{eq:TAB_bg}, a discussion of signs is in order.
Classically, the signs of the conserved current given in \eqref{jcurrent} are determined by the choice of metric signature; in particular, for a Euclidean metric all entries of $j^{AB}$ would be positive. 
The signature of our proposed effective metric, on the other hand, has so far not been fixed by the quantum theory. 
Given that $\sgn{\ev{\T^{aa}_0}} = - \sgn{\mkt}$ and $\sgn{\ev{\T^{00}_0}} = \sgn{\mkt}\sgn{\mb^2 - \ma^2}$, the initial conditions $\ma, \ \mb$ determine whether the effective metric we reconstruct is Euclidean or Lorentzian.
To fix the signature convention (i.e., ``East Coast'' or ``West Coast''), one could choose a preferred $\sgn{\mkt}$: identification with the classical discussion of section \ref{sec:current} would require $\sgn{\mkt}=1$. However, for $\sgn{\mkt}=-1$ 
we could also simply identify $j^{AB}$ with $-\ev{\T^{AB}}$, given that the symmetry arguments used to make this identification would be compatible with any constant rescaling, and already our definition (\ref{eq:TGFTdef}) of the energy-momentum tensor could equally well be replaced by a definition with the opposite sign.

In order to obtain a Lorentzian metric, we restrict to the case $\mb^2 > \ma^2$ and find the following effective expressions for the momentum of the clock field and the scale factor in the case of squeezing modes \eqref{eq:TAB_bg}:\footnote{In Euclidean signature all entries in \eqref{eq:jAB_bg} would have the same sign, and we would have to restrict to $\ma^2 > \mb^2$ instead; we would then have $\pi_0 =  |m||\mb^2 - \ma^2|$ and the identification $\frac{a^4}{\pi_0} = -\ev{\T^{aa}_0}$ would remain unchanged.}
\begin{align}
    \pi_0 = & \ev{\T^{00}_0} =  |m| (\mb^2 - \ma^2), \\
   a^4 = & -\pi_0\ev{\T^{aa}_0} =    m^2 (\mb^2 - \ma^2)\left( (\ma^2 + \mb^2)\cosh(2 |m| \chi^0) - 2 \ma \mb \sinh(2 |m| \chi^0) \right),
\label{eq:jTident}
\end{align}
from which we can calculate an effective Friedmann equation 
\begin{align}
    H^2 = \frac{1}{4} m^2 \left(1 - \frac{4(\ma^2 - \mb^2)^2}{((\ma - \mb)^2 e^{2 m \chi^0} + (\ma + \mb)^2 e^{-2 m \chi^0})^2} \right) = \frac{1}{4} m^2 \left(1 - \frac{\pi_0^4}{a^8} \right) \quad \underset{\rm late \ times}{\longrightarrow} \quad \frac{1}{4} m^2.
\label{eq:effFried}
\end{align}
We thus obtain a bounce at $a^4=\pi_0^2$, or equivalently, $\ev{\T^{aa}_0}^2 = \ev{\T^{00}_0}^2$.

The late-time limit -- $H^2$ going to a constant -- agrees with all Friedmann equations previously obtained for GFT models with a single clock (matter) field (see, e.g., \cite{Gielen_2020}) as well as with the case of general relativity with a single massless scalar field where $H^2=\frac{\kappa}{6}$, see (\ref{eq:clFried}).
This is an interesting observation given that in previous works effective Friedmann equations were always obtained by studying the evolution of the number operator $N = A^\dagger A$, whose dynamics for the squeezing case are given by $\ev{N} = (\ma^2 + \mb^2)\cosh(2 |m| \chi^0) - 2 \ma \mb \sinh(2 |m| \chi^0)$.
In the case of a single $J$ mode, the number operator is proportional to the GFT volume operator $V$, whose expectation value is identified with the classical volume element, hence in previous literature $\ev{N} \propto a^3$.  For our choice of state in the saddle-point approximation, the expectation value of the energy-momentum tensor component $\T^{aa}_0$, leading to \eqref{eq:jTident}, is in fact proportional to the expectation value of the number operator; hence our expression for $a^4$ is proportional to the one derived in previous literature for $a^3$. Within our simplifying assumptions this result would suggest that the physical picture of Planck-scale quanta of fixed volume whose total number is proportional to the total volume of space is not correct; rather, the relation $\ev{N}\propto a^4$ following from our new proposal suggests that the number of fundamental GFT degrees of freedom being excited grows more rapidly with the effective macroscopic scale factor $a$ than the classical volume which is proportional to $a^3$.

Notice that this mismatch in the power of $a$ identified with the particle number does not affect the agreement of the Friedmann equation with general relativity (with one massless scalar field) at late times; this agreement only requires that $H=(\log a)'$ should go to a constant, which is compatible with identifying $\ev{N}$ with any power of $a$.
(The GFT Friedmann equation could also have an additional $\frac{1}{\ev{N}}$ term as in \cite{Gielen:2020fgi},  which disappears here due to the normal ordering procedure we impose.) In our case, $m$ can be fixed to $m^2 = \frac{2}{3} \kappa$ to reproduce the Friedmann equation of general relativity with a single massless scalar field in the large volume limit, which is consistent with all previous literature. The only change is in the numerical factor needed in the identification of $m$ and $\kappa$. 

We saw, however, that the classical Friedmann equation for a model with four massless scalar coordinate fields \eqref{eq:clFried} differs from that with a single clock field. The additional contributions coming from spatial coordinate fields become dominant when $\frac{a^4}{\pi_0^2} \sim 1$, which is exactly when the bounce appears in (\ref{eq:effFried}). There is no early-time regime where these fields would not yet dominate classically, but we are still far away from the bounce in the GFT model. Hence, the GFT Friedmann equation does not match the Friedmann equation for a classical scenario with four massless scalar fields. We discuss the origin of this discrepancy and possible extensions that could lead to its alleviation in the conclusion.

Finally, one might be interested in the effective evolution of curvature through the bounce.
Consider the Ricci scalar, which for an FLRW spacetime for our lapse choice reads   $R = \frac{6 \pi_0^2}{a^6}(- 2 H^2 + \frac{a''}{a} )\,$. As the effective scale factor remains finite throughout, so does $R$ and the singularity is avoided. 
At the bounce, where we have $a' =0$ and $\pi_0 = a^2$, the value of the Ricci scalar is given by $R_{\rm bounce} = 6 \frac{a''}{a^3}$. Using $a''_{\rm bounce} = m^2 a_{\rm bounce}$ from \eqref{eq:jTident}, we find that the curvature at the bounce for our coherent state is given by $
R_{\rm bounce} = 6 \frac{m^2}{\pi_0} $ 
and is thus determined completely by the value of $\pi_0$ if $m$ is fixed by late-time consistency of the Friedmann equation.
\\

For an oscillating mode with expectation values of $\T^{AB}$ given in \eqref{eq:TAB_bg_osc}, we have the identification
\begin{align}
    \pi_0  = &\; \sgn{\mkt} |m| (\ma^2 + \mb^2)\, ,\\
    a^4 = & -|m|^2  (\ma^2 + \mb^2)\left( (\ma^2 - \mb^2) \cos(2 |m| \chi^0) - 2\, \sgn{\mkt}\ma \mb \sin(2 |m| \chi^0)\right)\,.
\end{align}
If we use the same convention as above with $\sgn{\mkt} = 1$, we again find a positive $\pi_0$, now for arbitrary initial conditions in terms of $ \ma$ and $\mb$.
The quantity $a^4$, however,  can change sign throughout the evolution due to its oscillatory behaviour. 
This is rather different to what one finds in the case with $a^3 \propto \ev{N}$, where the number operator remains constant for oscillating modes (with an expectation value that is always positive). It is clear that a single oscillating mode cannot lead to a realistic cosmology, however, if one includes both mode types, squeezing modes would lead to an expanding universe and oscillating modes would constitute an additional modulation in the evolution of the scale factor, whose relative effect is reduced with the expansion. 

Finally, we would like to point out that $\ev{\T^{00}_0}$ is time-independent already prior to the saddle-point approximation, implying that in this cosmological model $\pi_0$ is always constant. This is an important consistency check, given that constancy of $\pi_0$ corresponds to the Klein--Gordon equation for the classical FLRW model. We have of course shown above that the energy-momentum tensor is always conserved at the quantum level, so that we are guaranteed to obtain the exact Klein--Gordon equation $\partial_A j^{AB}=0$; however, what is nontrivial is to show that our quantum state can be consistently interpreted as an FLRW universe. This interpretation has been substantially strengthened in our new approach, 
given that the effective metric we recover explicitly gives a flat FLRW spacetime.

\section{Conclusion}\label{sec:conc}

In this article we proposed a novel set of operators for GFT which we used to reconstruct an effective metric directly from the quantum theory. This proposal goes beyond the entire previous GFT literature in which only a limited set of geometric observables, usually derived from the volume (or defining effective anisotropies \cite{Calcinari:2022iss}), were used.
Working in a (deparametrised) Hamiltonian setting, we established a relational coordinate system by coupling four massless scalar fields to the GFT action, where one of these fields is singled out as a clock field and the others serve as spatial coordinates. The GFT action remains unchanged upon translation of the scalar fields, leading to a conserved GFT energy-momentum tensor according to Noether's theorem. This translational symmetry represents the spacetime symmetry of constant shifts in the matter fields, which in that context leads to a conserved current directly related to the metric in the relational coordinate system. Hence, we proposed to identify the expectation values of the GFT energy-momentum tensor with components of the classically conserved current, obtaining an effective metric. 
We showed that the classical conservation law for the GFT energy-momentum tensor arising from the symmetry holds also at operator level, for different possible choices of operator ordering. In particular, this applies to the most relevant case of a normal-ordering prescription  which removes divergent contributions to the GFT energy-momentum tensor. In general, the free GFT action is decomposed into two different types of field modes, squeezed and oscillating modes, whose dynamics differ and which appear in various combinations in the expressions for the GFT energy-momentum tensor. 
The conservation law holds regardless of the choice of state, so that the matter fields always satisfy the classical Klein--Gordon equation exactly.

While our proposal is very  general, in the sense that an effective metric can be associated to any state that is sufficiently semiclassical to allow operator expectation values to be considered as effective quantities, the particular choice of state governs the specific form of such a metric and its symmetries.
We tested our proposition in a simple example: homogeneous cosmology as described by a flat FLRW metric. 
Here our choice of state was based on a few physical requirements. To ensure sufficient semiclassicality, we chose Fock coherent states as commonly done in the GFT literature. Secondly, the spacetime we are reconstructing should be spatially homogeneous, which is why the coherent state is peaked around the homogeneous $\vec{k}=0$ mode.
 
We found that the canonically conjugate momentum of the clock field is conserved in time, as it must be from the Klein--Gordon equation. Interestingly, only half of the possible space of initial conditions leads to an effective Lorentzian metric, with the other half leading to a Euclidean signature metric instead. This observation seems compatible with the fact that the basic assumptions we have made about the GFT model itself (choice of compact gauge group, neglecting interactions) would in principle be compatible with either Lorentzian or Euclidean models for quantum gravity. Moreover, the metric signature would be invisible if (as in almost all of the past work) one only has access to a Friedmann equation, which can take a very similar form in Euclidean or Lorentzian signature. (At the level of perturbations, the spacetime signature can of course be inferred from the equations of motion; here the work of \cite{Marchetti:2021gcv,Jercher:2023nxa} similarly finds that the effective signature depends on initial conditions, rather than being determined by choices such as whether one uses compact or non-compact gauge groups.) We also found an effective Friedmann equation that agrees with the previous GFT literature for models with a single massless scalar field; it agrees with general relativity (coupled to a single scalar field) at late times, and resolves the singularity by a bounce. We find a relation between the number operator and the effective scale factor that differs from the literature, namely $\ev{N} \propto a^4$ instead of $\ev{N} \propto a^3$. 
Oscillating modes would give a contribution to the effective scale factor for which $a^4$ can take negative values. Therefore, it seems that such modes can only appear in conjunction with one or more squeezed modes. 

While we find agreement with the standard GFT scenario for homogeneous cosmology, the agreement with the classical Friedmann equation is not entirely satisfactory: given that the spatial coordinate fields need to have nonvanishing gradient energy, they would be expected to contribute additional terms in the energy density that can only be neglected when $\frac{\pi_0^2}{a^4} \gg 1$. Such terms are not seen in the effective GFT dynamics, so that at best one might expect a matching between GFT and classical cosmology for early times, where these terms do not yet dominate. However, we found that the bounce already occurs at $\frac{\pi_0^2}{a^4} = 1$, so that there is no such early-time regime in the GFT setting. The bounce scale is determined by initial conditions, namely the value of $\pi_0$, and unrelated to quantities like the ratio of the energy density to the Planck density. This seems to imply that the bounce could happen at low curvatures (somewhat reminiscent of what happens in the so-called $\mu_0$-scheme of loop quantum cosmology \cite{Ashtekar:2011ni}), which needs to be clarified further. 

One might think that the two Friedmann equations can be brought into agreement if one deviates from the assumption of a flat FLRW universe on the classical side; indeed, a positive curvature term could cancel the contribution from the spatial matter fields. In previous work on GFT cosmology, where one only had access to the Friedmann equation, such an interpretation would have been viable. However, in our scenario we have access to all metric components, and since $\ev{\T^{ab}_0}\propto \delta^{ab}$, our choice of state clearly corresponds to a flat metric. The mismatch seems to arise from a general difficulty to include spatial gradients of the scalar fields in the GFT construction. As we have seen, spatial homogeneity means that the mean field should be peaked around $\vec{k}=0$; indeed we have neglected all finite $\vec{k}$ contributions in our saddle-point approximation. These assumptions then also imply that the canonical momenta (or, in other words, the kinetic energy) of the spatial coordinate fields must vanish. It would not be possible to introduce non-vanishing canonical momenta without also departing from homogeneity, as discussed from a slightly different perspective in \cite{Gielen:2020fgi}. In our classical setup we had to assume that while the spatial coordinate fields are not homogeneous in space (they would not be good coordinates otherwise), their energy-momentum tensor is. It may be that in our GFT scenario the imposition of spatial homogeneity by peaking on $\vec{k}=0$ actually imposes a stronger condition of homogeneity on the matter fields themselves, which would be incompatible with the presence of gradient energy, and imply an inconsistency in our starting point of assuming that the matter fields are good spatial coordinates. Our results here are consistent with previous literature, in which the spatial coordinate fields are either simply assumed to be negligible at background level \cite{Marchetti:2021gcv,Jercher:2023nxa}, or where effective Friedmann dynamics only show contributions from additional matter fields if one chooses a state peaked around $\vec{k}_0\neq 0$ \cite{Gielen:2020fgi}. Introducing gradient energy into the effective Friedmann equation might require entirely different types of states, for instance states built from multiple Peter--Weyl modes. It might also require including the effects of GFT interactions, which we neglected here; as stated above, this assumption usually applies to the early universe since interactions generally dominate at late times. 
If the late-time limit of a suitably defined interacting GFT matches with the expression for general relativity with four scalar fields, this could be seen as a phenomenological constraint on the allowed types of GFT interactions.

A possible alternative to the model we studied would be to introduce an additional scalar matter field as in \cite{Marchetti:2021gcv,Jercher:2023nxa}. This fifth field is not interpreted as a relational coordinate and has its own independent initial conditions; if this field dominates over the coordinate fields at some initial time, such a scenario might yield an intermediate regime where the effective Friedmann equation matches that of general relativity, before spatial gradient terms would be expected to dominate. Whether this is realised in our new approach needs to be studied in more detail.
Another immediate question for future work would be how different types of coherent state, such as those with a mean field peaked around $\vec{k}_0\neq 0$, can be interpreted in terms of the effective metric introduced in this paper. More generally, the constructions shown here can be extended to small perturbations of an FLRW universe or to situations such as spherical symmetry, where they could describe the dynamics of an effective GFT black hole.

\acknowledgments
This work was supported by the Royal Society through the University Research Fellowship
Renewal URF$\backslash$R$\backslash$221005 (SG).

\appendix

\section{Saddle-point approximation: Next-to-leading order}
\label{app:saddlePointNLO}

So far we have only considered the leading-order term in the saddle-point approximation, effectively recovering the scenario of an infinitely peaked Gaussian, where only the $\Vec{k}=0$ mode contributes to the effective dynamics. However, as pointed out previously, in our construction this limit cannot be realised exactly and inhomogeneous modes will inevitably contribute to what is usually understood as background dynamics. 
In what follows we therefore include the next-to-leading-order term in the saddle-point approximation and consider its implications for the effective dynamics. We also establish the latest time at which the saddle-point approximation can be valid. 

Including the next-to-leading-order term in \eqref{eq:saddlePtApproxLO}, the saddle-point approximation reads
\begin{align}
     \int_a^b \dd^3 x\, g(\Vec{x}) e^{- \lambda  (\Vec{x}-\Vec{\mu})^2} \approx \sqrt{\frac{\pi}{\lambda}}^3 \left(g(\Vec{\mu}) + \frac{1}{4 \lambda} \Laplace g(\Vec{x})|_{\Vec{x} = \Vec{\mu}}\right) \, ,
     \label{eq:3dSaddleP}
\end{align}
where $\lambda  = \frac{1}{s^2} > 0$ and $g(\Vec{x})$ is integrable, i.e., $\int_a^b |g(\Vec{x})| \dd^3 x < \infty$.\footnote{Note that strictly speaking the integrals we consider are over the range $(-\infty, \infty)$, with $g(\Vec{x})$ increasing monotonically with $\Vec{x}$. We then need to limit integration to a finite range, which is tantamount to excluding modes with very large wave numbers.} The saddle point approximation becomes increasingly accurate for $\lambda \to \infty$. 
In our case, we have $\vec\mu=0$. 
For a derivation of the saddle-point approximation and its higher-order terms, see \cite{miller2006applied}.

We apply the above to the components of the GFT energy-momentum tensor $\langle \sigma | \T^{AB} | \sigma \rangle$, with $\sigma$ specified in \eqref{eq:sigmaGauss} and the $\T^{AB}$ operators given in \eqref{eq:TABoperatorsNormal}. 
Recalling that the off-diagonal components $\ev{\T^{0b}}$ and $\ev{\T^{a\neq b}}$ are exactly zero, independent of the saddle-point approximation,
we  need to consider only $\ev{\T^{00}_0}$ and $\ev{\T^{aa}_0}$.

For $\ev{\T^{00}}$, we have $g(\Vec{\gamma}) = |\omega_\gamma| = \sqrt{\Vec{\gamma}^2 + m^2}$ and thus $\Laplace g(\Vec{\gamma})|_{\Vec{\gamma} = 0} = \frac{3}{|m|}$.
If we include the next-to-leading-order correction to the expression in \eqref{eq:TAB_bg}, we obtain a small constant shift in the value of $|\pi_0| = \ev{\T^{00}_0}$, namely
\begin{align}
    \ev{\T^{00}_0}_{\rm NLO}  &     \approx \sgn{\mkt} \left(|m| + \frac{3}{4 |m| \lambda}\right) (\mb^2 - \ma^2) \,. 
\end{align}

For the $\ev{\T^{aa}}$ component, on the other hand, we have more complex expressions that include also a time dependence (see \eqref{eq:TAB_bg}): 
\begin{align}
\begin{split}
    g(\Vec{\gamma}) = &    \left(- |\omega_\gamma| + \frac{\gamma_a^2}{|\omega_\gamma|}\right) \left( (\ma^2 + \mb^2)\cosh(2 |\omega_\gamma| \chi^0) - 2\, \sgn{\mkt} \ma \mb  \sinh(2 |\omega_\gamma| \chi^0) \right)\\
    & \, + \frac{\gamma^2_a}{|\omega_\gamma|} (\ma^2 - \mb^2)\, , \\
    \Laplace g(\Vec{\gamma})|_{\Vec{\gamma} = 0} = & 
    -\frac{1}{|m|} \left(\left(\ma^2+\mb^2\right) \cosh (2 |m| \chi^0)-2 \ma \mb \sgn{\mkt} \sinh (2 |m| \chi^0)\right)\\
    & -6\chi^0  \left( \left(\ma^2+\mb^2\right) \sinh (2 |m|
   \chi^0)-2 \ma \mb \sgn{\mkt} \cosh (2 |m| \chi^0)\right)+\frac{2}{|m|} \left(\ma^2-\mb^2\right)\,. 
\end{split}
\label{eq:gTaa}
\end{align}
As $\lambda$ is large, but finite, $g(\Vec{\gamma})$ as given above will dominate the integral \eqref{eq:3dSaddleP} for late enough times, leading to an inapplicability of the saddle-point approximation. To establish the maximum value of $\chi^0$ for which the saddle-point approximation is viable, we consider the late-time limit where  $g(\Vec{\gamma}) \propto e^{2 \omega_\gamma \chi^0}$.
The integrand then behaves as $e^{-\lambda \Vec{\gamma}^2} g(\Vec{\gamma}) \propto e^{- \lambda \Vec{\gamma}^2 + 2 \omega_\gamma \chi^0}$. The saddle-point approximation is applicable at a maximum of $- \lambda \Vec{\gamma}^2 + 2 \omega_\gamma \chi^0$, which occurs at $\Vec{\gamma} = 0$ for $\chi^0 <  |m| \lambda$. 
Additionally, the condition $g(0) > \frac{1}{4 \lambda} \Laplace g(\Vec{\gamma})|_{\Vec{\gamma} = 0}$ needs to be satisfied to ensure that $g(0)$ remains  the leading-order term throughout the evolution. 
In the late-time limit, \eqref{eq:gTaa} translates to
\begin{align}
\begin{split}
    g(0) \approx &    - \frac{|m|}{2} \left( \ma - \, \sgn{\mkt} \mb  \right)^2 e^{2 |m| \chi^0}\, , \\
    \Laplace g(\Vec{\gamma})|_{\Vec{\gamma} = 0} \approx & 
    -  \left( \ma - \, \sgn{\mkt} \mb  \right)^2 e^{2 |m| \chi^0} \left(\frac{1}{2 |m|} + 3 \chi^0 \right) \, ,\\
    %
   \Rightarrow\quad  g(0) > & \frac{1}{4 \lambda} \Laplace g(\Vec{\gamma})|_{\Vec{\gamma} = 0} \qquad \Leftrightarrow 
    \qquad \chi^0 \, < \, \frac{2}{3} \lambda |m| - \frac{1}{6 |m|}\, ,
\end{split}
\end{align}
which gives a similar, but more stringent constraint on the latest time the saddle-point approximation can be considered as valid. 

Restricting to this regime, the expression for $\ev{\T^{aa}_0}$ then reads\\
\begin{align}
\begin{split}
     \ev{\T^{aa}_0}_{\rm NLO} \approx &  \,\sgn{\mkt}  
     \Big[ \frac{\ma^2 - \mb^2}{2 \lambda |m|} + \left(|m|+\frac{1}{4 \lambda 
   |m|}\right) \left(2\ma \mb \sgn{\mkt} \sinh (2 |m| \chi^0) -\left(\ma^2+\mb^2\right) \cosh (2 |m| \chi^0)\right)\\
   & \ +\frac{3 \chi^0}{2\lambda} \left(2 \ma \mb \sgn{\mkt} \cosh (2 |m| \chi^0)- \left(\ma^2+\mb^2\right) \sinh (2 |m| \chi^0)\right) \Big] \, .
\end{split}
   \label{eq:TaaNLO}
\end{align}
If we recall that the above is related to the scale factor as $a^4 = - |\pi_0| \ev{\T^{aa}_0} $, it is apparent that the additional terms will influence the form of the effective Friedmann equation, as we have
\begin{align}
    \frac{a'}{a} = \frac{1}{4} \frac{\ev{\T^{aa}_0}'_{\rm NLO}}{ \ev{\T^{aa}_0}_{\rm NLO}} = \frac{1}{4} \frac{ g'(0) + \frac{1}{4 \lambda} \Laplace g'(\Vec{\gamma})|_{\Vec{\gamma} = 0}}{g(0) + \frac{1}{4 \lambda} \Laplace g(\Vec{\gamma})|_{\Vec{\gamma} = 0}}\, .
    \label{eq:FriedNLO}
\end{align}

We first focus our attention on the implications of the next-to-leading-order terms in the late-time limit, in which the  expectation value \eqref{eq:TaaNLO} and its logarithmic derivative \eqref{eq:FriedNLO} simplify to
\begin{align}
   \ev{\T^{aa}_0}_{\rm NLO} & \approx -\sgn{\mkt}e^{2 |m| \chi^0} \left(\ma- \sgn{\mkt} \mb\right)^2 \left(\frac{|m|}{2} + \frac{3\chi^0}{4\lambda}+\frac{1}{8\lambda|m|}\right)  \,, \label{eq:TaaNLOlate}\\
     \left(\frac{\ev{\T^{aa}_0}'}{ \ev{\T^{aa}_0}}\right)_{\rm NLO}^2 &  \approx \frac{16 m^2 \left(2 \lambda  m^2+3 |m| \chi^0+2\right)^2}{\left(4 \lambda  m^2+6 |m| \chi^0+1\right)^2} \,, \label{eq:FriedNLOlate}
\end{align}
where we neglected the subdominant constant term in \eqref{eq:TaaNLO}. 

For large values of $\chi^0$ we have $\left(\ev{\T^{aa}_0}'/ \ev{\T^{aa}_0}\right)_{\rm NLO}^2 \to 4 m^2$ such that $H^2_{\rm NLO} \to \frac{m^2}{4}$, as in the main text. Note however that at the end of the period of validity for the saddle-point approximation, one has $H^2_{\rm NLO} = \frac{1}{16} (2 m+ \frac{3}{4 \lambda  m})^2$, which is slightly larger than for the leading-order contribution only. 

\begin{figure}
     \centering
     \begin{subfigure}[t]{0.45\textwidth}
         \centering
         \includegraphics[width=\textwidth]{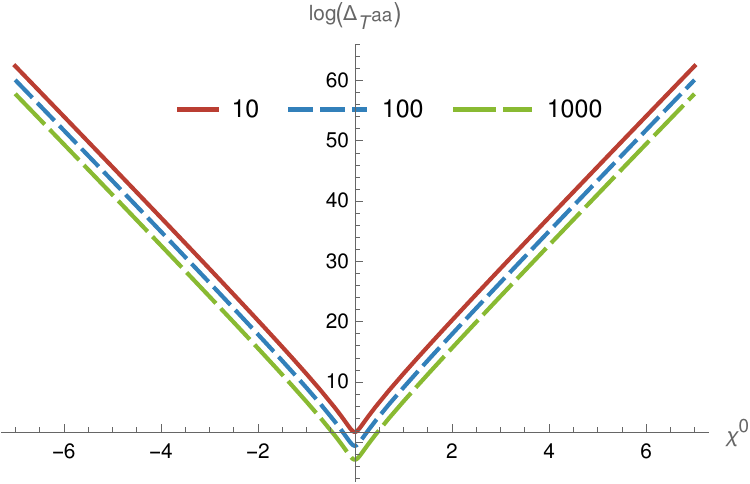}
         \caption{$\ln(|\ev{\T^{aa}_0}_{\rm NLO}| - |\ev{\T^{aa}_0}_{\rm LO} |)$. As $\mb > \ma$, it is clear from \eqref{eq:TaaNLO} that the next-to-leading-order expression is greater than the leading-order expression alone.}
         \label{fig:y equals x}
     \end{subfigure}
     \hfill
     \begin{subfigure}[t]{0.45\textwidth}
         \centering
         \includegraphics[width=\textwidth]{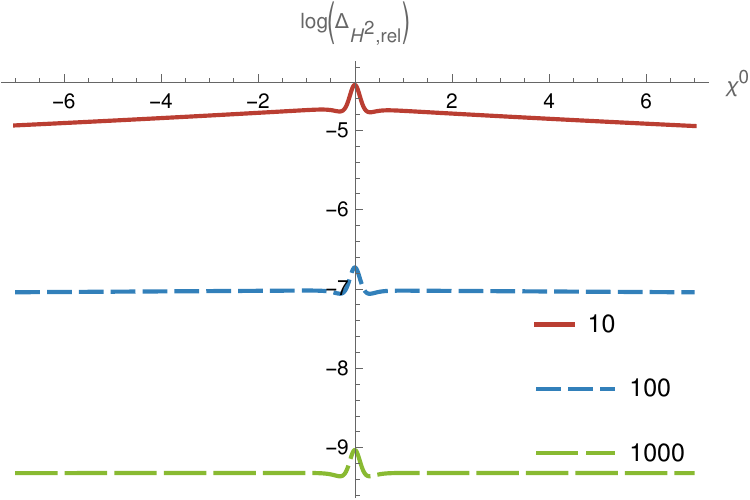}
         \caption{Relative difference of $H^2$, as given in \eqref{eq:relativeError}. The relative difference is largest around the bounce. Taking the limit $\chi^0 \to 0$ gives a finite value, as depicted above (e.g., $ \Delta_{H^2, \, \rm rel} \approx 0.002$ for $\lambda = 10$).}
         \label{fig:three sin x}
     \end{subfigure}
        \caption{Comparison of dynamics arising from including also the next-to-leading-order term, $\ev{\T^{aa}_0}_{\rm NLO}$, to those with the leading-order only, $\ev{\T^{aa}_0}_{\rm LO}$, for different values of $\lambda$, namely $\lambda = 10$ (red, full) , $\lambda = 100$ (blue, small dashed), $\lambda = 1000$ (green, large dashed). The other parameter values are $|m| = 4\sqrt{\pi/3}$, $\ma = 10$, $\mb = 20$, $\sgn{\mkt} = 1$. The times were adjusted such that the bounces coincide and occur at $\chi^0=0$.}
        \label{fig:relativeErrorlambda}
\end{figure}

To establish the effect of the next-to-leading-order term on the bounce behaviour we consider the absolute difference between $\ev{\T^{aa}_0}_{\rm NLO}$ and $\ev{\T^{aa}_0}_{\rm LO}$ as well as the relative difference of the leading-order and next-to-leading-order Hubble rate, defined as
\begin{align}
   \Delta_{H^2, \, \rm rel} := \left( \left(\frac{\ev{\T^{aa}_0}'}{ \ev{\T^{aa}_0}}\right)_{\rm NLO}^2 - \left(\frac{\ev{\T^{aa}_0}'}{ \ev{\T^{aa}_0}}\right)_{\rm LO}^2\right) \left(\frac{\ev{\T^{aa}_0}'}{ \ev{\T^{aa}_0}}\right)_{\rm LO}^{-2}\, ,
    \label{eq:relativeError}
\end{align}
where we correct for different bounce times, such that both bounces occur at $\chi^0 = 0$.\footnote{For the leading-order term only, the bounce happens at $\chi^0 = \frac{1}{2m}\ln\left(\frac{|\ma+ \sgn{\mkt} \mb|}{|\ma+ \sgn{\mkt} \mb|} \right)$.}
Both are depicted for different values of $\lambda$ in fig.\,\ref{fig:relativeErrorlambda}. 
The qualitative bounce behaviour remains unchanged by the contribution of inhomogeneous modes. The relative error peaks around the bounce, where $H^2$ is minimal, and decreases with increasing $\chi^0$, as can also be seen from \eqref{eq:FriedNLO}. 
Overall, next-to-leading-order contributions have minimal impact on the dynamics, demonstrating that they can be neglected as is done in the main text.

\bibliographystyle{JHEP}

\bibliography{bib2}

\end{document}